\documentclass[a4paper,11pt]{article}
\pdfoutput=1 

\usepackage{jheppub} 

\usepackage[T1]{fontenc} 
\usepackage[english]{babel}
\usepackage[utf8]{inputenc}
\usepackage{cancel}
\usepackage{physics}
\usepackage[dvipsnames]{xcolor}
\usepackage{booktabs}
\usepackage{fontawesome}
\usepackage{subcaption}

\newcommand{\CB}{\mathrm{CB}}
\newcommand{\CP}{\cancel{\mathrm{CP}}}
\newcommand{\EW}{\mathrm{EW}}
\newcommand{\eff}{\mathrm{eff}}
\newcommand{\tree}{\mathrm{tree}}
\newcommand{\cl}{\mathrm{cl}}

\newcommand\myshade{80}
\colorlet{mylinkcolor}{ForestGreen}
\colorlet{mycitecolor}{Red}
\colorlet{myurlcolor}{violet}
\hypersetup{
  linkcolor  = mylinkcolor!\myshade!black,
  citecolor  = mycitecolor!\myshade!black,
  urlcolor   = myurlcolor!\myshade!black,
  colorlinks = true
}

\usepackage[frozencache,cachedir=.]{minted}
\newcommand{\mil}[1]{\mintinline{julia}{#1}}

\title{One-Loop Charge-Breaking Minima in the Two-Higgs Doublet Model}
\author[a,b]{P.M. Ferreira,}
\author[c,d]{L. A. Morrison,}
\author[c,d]{S. Profumo}

\affiliation[a]{Instituto Superior de Engenharia de 
Lisboa~---~ISEL,\\1959-007 Lisboa, Portugal}
\affiliation[b]{Centro de F\'{\i}sica Te\'{o}rica e 
Computacional~---~CFTC,\\
FCUL, 1749-016 Lisboa, Portugal}
\affiliation[c]{Department of Physics, 1156 High St., University of California Santa Cruz, Santa Cruz, CA 95064, USA}
\affiliation[d]{Santa Cruz Institute for Particle Physics, 1156 High St., Santa Cruz, CA 95064, USA}

\emailAdd{pmmferreira@fc.ul.pt}
\emailAdd{loanmorr@ucsc.edu}
\emailAdd{profumo@ucsc.edu}

\abstract{We analyze the vacuum structure of the one-loop effective potential in the two Higgs doublet model. We find that electroweak-breaking vacuua can coexist with charge breaking ones, contradicting a theorem valid at tree-level. We perform a numerical analysis of the model and supply explicit parameter values for which charge-breaking vacuua can be the global minimum of the theory, and deeper than charge-preserving ones.}

\begin{document} 
\maketitle
\flushbottom

\section{Introduction}
\label{sec:intro}

The discovery by the LHC collaborations of the Higgs boson  \cite{Aad:2012tfa,Chatrchyan:2012xdj} provided the missing piece of the puzzle for the Standard Model (SM) of particle physics. Since then, measurements of the Higgs' properties \cite{Khachatryan:2016vau} have shown that this scalar, with mass around 125 GeV, behaves
largely as expected in the minimal SM: thus far, and within the measured precision, no significant deviations from SM-like behavior have been observed. But the SM leaves a great many questions unanswered, such as the origin of the matter-antimatter asymmetry, the nature of dark matter as a particle, the observed fermion mass hierarchy, and the strong CP problem. SM extensions are therefore of interest to attempt to provide answers to these, and other, unsolved problems. Models with extended scalar sectors, in particular, are quite popular and widely studied in the literature. One of the simplest beyond the SM theories is the two Higgs doublet model (2HDM), first proposed by Lee in 1973~\cite{Lee:1973iz} to provide an additional source of CP violation stemming from the scalar sector through spontaneous symmetry breaking.

In the 2HDM, the gauge and fermion content are the same as in the SM, but instead of a single $\mathrm{SU}(2)$ doublet with hypercharge $Y =  1$,we now have two, $\Phi_1$ and $\Phi_2$. This leads to a rich phenomenology (see~\cite{Branco:2011iw} for a review), boasting a richer scalar spectrum than the SM's, with two CP-even scalars, a pseudo-scalar and a charged scalar. The model can have tree-level flavor-changing neutral currents mediated by scalars, can provide a dark matter candidate and have spontaneous CP violation. The 2HDM easily reproduces all experimental results from the SM, and indeed it has a {\em decoupling limit} where the extra scalars are very massive and the model's predictions can be made to be virtually indistinguishable from those of the SM. The 2HDM also has a richer vacuum structure then the SM -- whereas in the SM the only possible vacuum is the one which breaks electroweak symmetry, in the 2HDM spontaneous CP breaking is also possible, as well as minima where the electromagnetic symmetry $U(1)_{em}$ is broken. These latter minima are unwanted since they would imply a massive photon. The possibility of charge (and color) breaking already arises in SUSY models, and leads to bounds on some of the parameters of the model~\cite{Frere:1983ag}.

The possibility of reducing the 2HDM parameter space in a similar manner -- imposing bounds on the model's parameters to avoid global charge-breaking (CB) minima -- is very appealing. In many cases, {\em sufficient} conditions to avoid CB minima were considered~\cite{Lee:1973iz,Velhinho:1994np,Gunion:2002zf}, but at the time it wasn't known whether such conditions were too restrictive. However,  in~\cite{Ferreira:2004yd,Barroso:2005sm} a remarkable result was obtained: the structure of the tree-level 2HDM scalar potential is such that, if an electroweak breaking minimum exists, any CB extremum that might then occur is necessarily a saddle point lying {\em above} the minimum. Likewise, it was shown that if a CB minimum exists, any stationary point which would break the normal electroweak symmetries is a saddle point lying above it. Analogous results were also proved for the relationship between electroweak extrema and CP breaking ones. Thus a 2HDM electroweak breaking minimum, if it exists, is guaranteed, {\em at tree level}, to be stable against tunneling to deeper CB or CP breaking vacuua, since such deeper minima were shown to not exist. This result was further studied in refs.~\cite{Maniatis:2006fs,Nishi:2006tg,Ivanov:2006yq,Ivanov:2007de,Maniatis:2007vn}. In particular, using a Minkowski formalism to rewrite the 2HDM scalar potential, Ivanov was able to show~\cite{Ivanov:2006yq,Ivanov:2007de} the stability of the different vacuua through geometric arguments. Other results concerning neutral minima in the 2HDM  were also obtained -- it was shown~\cite{Ivanov:2006yq,Ivanov:2007de,Barroso:2007rr} that neutral minima can coexist in the 2HDM scalar potential, provided they break the same symmetries. This had implications for the Inert model~\cite{Deshpande:1977rw,Barbieri:2006dq,Cao:2007rm}, a version of the 2HDM where a discrete $\mathbb{Z}_2$ symmetry is preserved by both the Lagrangian and by the vacuum -- the vacuum preserves $\mathbb{Z}_2$ since only one of the doublets acquires a non-zero vacuum expectation value. Two possibilities for minima then arise, depending on which of the doublets has the non-zero VEV, the Inert Minimum (where fermions acquire mass after spontaneous symmetry breaking) and the Inert-Like Minimum (where the fermions remain massless). In~\cite{Barroso:2007rr,Ivanov:2007de} expressions relating the depth of the potential at each of these minima were found, and it was shown that, for specific regions of parameter space, they could coexist.

However, powerful though the demonstrations of~\cite{Ferreira:2004yd,Barroso:2005sm} and~\cite{Ivanov:2006yq,Ivanov:2007de} were, those works dealt with the tree-level potential. The expressions found there comparing the depth of the potential at different extrema depended heavily on tree-level formulae
for the scalar masses; for the minimization conditions determining the vacuum expectation values (VEVs); and for the potential itself. A valid question is therefore whether these results are robust when one considers loop  corrections to the potential -- will the stability theorems deduced for the tree-level potential still hold at one-loop? The first hint that that may not be the case was obtained in~\cite{Ferreira:2015pfi}, where a one-loop calculation was undertaken to analyze the coexistence of neutral minima in the Inert model. The effective potential formalism was employed and it was shown that, in certain cases, a tree-level local minimum could become a one-loop global one, and vice-versa. This (rare) possibility occurred only for regions of parameter space where the tree-level minima were close to degenerate, hence it did not correspond to a breakdown in perturbation theory, rather it implied loop corrections could change the nature of tree-level vacuua. Further, the one-loop calculation enlarged the region of parameter space for which different neutral minima could coexist.

The purpose of this paper is to investigate, using the one-loop effective potential, whether the conclusions concerning the (non-)coexistence of neutral and charge breaking vacuua in the 2HDM hold when radiative corrections are taken into account. We review the tree-level results for the classical 2HDM potential in section~\ref{sec:tree} then proceed to review the formalism of the one-loop effective potential in section~\ref{sec:one_loop}, including a discussion of issues related with gauge fixing. The numerical methods we  use to carry the minimization of the one-loop potential are detailed in section~\ref{sec:num} where we also present results of numerical scans of the model's parameter space and give a few illuminating examples. We draw our conclusions in section~\ref{sec:conc}.

\section{The tree-level vacuum structure of the 2HDM}\label{sec:tree}

The 2HDM contains two hypercharge 1 $SU(2)$ scalar doublets, and the most general scalar potential one can write has a total of 14 real parameters. Since both doublets are identical, any linear combination of them which preserves the scalar kinetic terms should lead to the same physics. This {\em basis invariance}, which corresponds to a redefinition of the fields via a $2\times 2$ unitary matrix $U$~\footnote{Meaning, the theory is physically equivalent if one considers the new doublets $\Phi_i^\prime = U_{ij} \Phi_j$.}, allows one to reduce the number of free parameters to 11~\cite{Davidson:2005cw}. This most general 2HDM will include flavor changing neutral currents (FCNC), mediated by neutral scalars at tree-level, when one considers the full Lagrangian, including fermions. To prevent this, a discrete $\mathbb{Z}_2$ symmetry is introduced~\cite{Glashow:1976nt,Paschos:1976ay}, such that $\Phi_1\rightarrow \Phi_1,\,\Phi_2\rightarrow -\Phi_2$, which is extended to the Yukawa sector in such a way that each class of same-charge fermions (up and down-type quarks and charged leptons) only couple to one of the doublets. This eliminates tree-level FCNC and, due to the several possibilities of extending $\mathbb{Z}_2$  to the Yukawa sector leads to four types of 2HDMs (type I, type II, lepton specific and flipped~\cite{Branco:2011iw}). So that the model can possess a {\em decoupling limit}~\cite{Gunion:2002zf}, a softly $\mathbb{Z}_2$ breaking quadratic term, $m_{12}^2$, is usually introduced, so that the scalar potential is characterized by 8 real independent parameters.

\subsection{Classical Potential} 

The 2HDM scalar potential we will be studying possesses a softly broken \(\mathbb{Z}_{2}\) symmetry and is therefore given, at tree-level, by
\begin{align}\label{eqn:vtree}
    V^{(0)}(\Phi) &= 
    m_{11}^{2}|\Phi_{1}|^{2} 
    + m_{22}^{2}|\Phi_{2}|^{2} 
    - m_{12}^{2}\left[\Phi_{1}^{\dagger}\Phi_{2} + \mathrm{h.c.}\right]\\
    & 
    +\dfrac{1}{2}\lambda_{1}|\Phi_{1}|^{4} 
    +\dfrac{1}{2}\lambda_{2}|\Phi_{2}|^{4}
    +\lambda_{3}|\Phi_{1}|^{2}|\Phi_{2}|^{2} 
    +\lambda_{4}|\Phi_{1}^{\dagger}\Phi_{2}|^{2}
    + \dfrac{1}{2}\lambda_{5}\left[\left(\Phi_{1}^{\dagger}\Phi_{2}\right)^{2} + \mathrm{h.c.}\right]\notag
\end{align}
where all the parameters are taken to be real~\footnote{By real we mean the CP symmetry that the unbroken $\mathbb{Z}_{2}$ potential had is left unbroken. Considering a complex coefficient $m^2_{12}$ would lead to a model with explicitly broken CP, known as the Complex 2HDM~\cite{Pilaftsis:1999qt,Ginzburg:2002wt,Khater:2003wq,ElKaffas:2007rq,ElKaffas:2006gdt,WahabElKaffas:2007xd,Osland:2008aw,Grzadkowski:2009iz,Arhrib:2010ju,Barroso:2012wz}. Further promoting the $\mathbb{Z}_{2}$ symmetry to a continuous $U(1)$ but keeping the complex soft breaking term and allowing for the possibility of flavor violation in the quark sector yields models with interesting phenomenology~\cite{Ferreira:2011xc,Ferreira:2019aps}, but not the subject of the current paper.}. The scalar doublets, \(\Phi_{1}\) and \(\Phi_{2}\) contain a combined eight real component fields which can be parameterized as follows:
\begin{align}\label{eqn:phi1_and_phi2}
    \Phi_{1} &= \dfrac{1}{\sqrt{2}}
    \begin{pmatrix}
    c_{1} + \mathrm{i}\,  c_{2}\\
    r_{1} + \mathrm{i}\, i_{1}
    \end{pmatrix}, &
    \Phi_{2} &= \dfrac{1}{\sqrt{2}}
    \begin{pmatrix}
    c_{3} + \mathrm{i}\, c_{4}\\
    r_{2} + \mathrm{i}\, i_{2}
    \end{pmatrix}\, .
\end{align}
It is well know that the 2HDM classical potential exhibits three-different types of extrema (for instance, see section 5.8 of~\cite{Branco:2011iw} for a demonstration): a \(U(1)_{\mathrm{EM}}\) and \(\mathrm{CP}\)-conserving extremum, which we denote by \(\expval{\Phi_{i}}_{\EW}\)
\begin{align}
    \expval{\Phi_{1}}_{\EW} &= \dfrac{1}{\sqrt{2}}
    \begin{pmatrix}
    0\\
    v_{1}
    \end{pmatrix}, &
    \expval{\Phi_{2}}_{\EW} &= \dfrac{1}{\sqrt{2}}
    \begin{pmatrix}
    0\\
    v_{2}
    \end{pmatrix}\, ,
\end{align}
a \(U(1)_{\mathrm{EM}}\)-violating extremum,
\begin{align}
    \expval{\Phi_{1}}_{\CB} &= \dfrac{1}{\sqrt{2}}
    \begin{pmatrix}
    \alpha\\
    \bar{v}_{1}
    \end{pmatrix}, &
    \expval{\Phi_{2}}_{\CB} &= \dfrac{1}{\sqrt{2}}
    \begin{pmatrix}
    0\\
    \bar{v}_{2}
    \end{pmatrix}
\end{align}
with the vev \(\alpha \neq 0\) breaking electric charge conservation and consequently giving a mass to the photon and a \(\mathrm{CP}\)-violating extremum:
\begin{align}
    \expval{\Phi_{1}}_{\CP} &= \dfrac{1}{\sqrt{2}}
    \begin{pmatrix}
    0\\
    v'_{1} + i\delta
    \end{pmatrix}, &
    \expval{\Phi_{2}}_{\CP} &= \dfrac{1}{\sqrt{2}}
    \begin{pmatrix}
    0\\
    v'_{2}
    \end{pmatrix}
\end{align}
where \(\delta \neq 0\). In this work, we will be focusing on the \(\EW\) and \(\CB\) extrema. 

\subsection{Classical Extrema} 

To investigate the relative depths of the classical potential evaluated at the \(\EW\) and \(\CB\) extrema, it is useful to introduce the following gauge-invariant variables~\cite{Velhinho:1994np,Ferreira:2004yd}:
\begin{align}
    x_{1} &= \left|\Phi_{1}\right|^{2} = c_{1}^{2} + c_{2}^{2} + r_{1}^{2} + i_{1}^{2}\\
    x_{2} &= \left|\Phi_{2}\right|^{2} = c_{3}^{2} + c_{4}^{2} + r_{2}^{2} + i_{2}^{2}\\
    x_{3} &= \Re\left(\Phi_{1}^{\dagger}\Phi_{2}\right) = c_{1}c_{3} + c_{2}c_{4} + i_{1}i_{2} + r_{1}r_{2}\\
    x_{4} &= \Im\left(\Phi_{1}^{\dagger}\Phi_{2}\right) = c_{1} c_{4} - c_{2} c_{3} + i_{2} r_{1} - i_{1} r_{2}
\end{align}
In terms of these variables, then, the classical potential of Eqn.~\eqref{eqn:vtree} is written as 
\begin{align}
    V^{(0)}(x_{1},x_{2},x_{3},x_{4}) &= \sum_{i=1}^{4}a_{i}x_{i} + \dfrac{1}{2}\sum_{i,j=1}^{4}b_{ij}x_{i}x_{j}
\end{align}
with real parameters $a_i$ and \(b_{ij} = b_{ji}\). In terms of the original parameters of Eqn.~\eqref{eqn:vtree}, the $a_i$ and $b_{ij}$ are given by
\begin{align}
    a_{1} &= m_{11}^{2}, & a_{2} &= m_{22}^{2}, & a_{3} &= -2m_{12}^{2}\\
    b_{11} &= \dfrac{\lambda_{1}}{2}, &
    b_{22} &= \dfrac{\lambda_{2}}{2}, &
    b_{33} &= \lambda_{4} + \lambda_{5}\\
    b_{44} &= \lambda_{4} - \lambda_{5}, &
    b_{12} &= \dfrac{\lambda_{3}}{2}
\end{align}
with all the unspecified parameters equal to zero. Collecting the \(x_{i}\)'s, \(a_{i}\)'s and \(b_{ij}\)'s in vectors \(X, A\) and symmetric matrix \(B\), respectively, we can rewrite the classical potential as
\begin{align}
    V^{(0)} = A^{T} X + \dfrac{1}{2}X^{T}BX\,.
\end{align}
The values of the vector $X$ at the \(\EW\) and \(\CB\) extrema are given by
\begin{align}
    X_{\EW}^T &= (v_{1}^{2}, v_{2}^{2}, v_{1}v_{2}, 0)\\
    X_{\CB}^T &= (\bar{v}^{2}_{1}, \bar{v}^{2}_{2} + \alpha^{2}, \bar{v}_{1}\bar{v}_{2}, 0)\,.
\end{align}
We can then see that  the non-trivial minimisation conditions of the classical potential at the $\EW$ extremum may be expressed in terms of \(X_{\EW}\) as
\begin{align}
    \dfrac{\partial V}{\partial r_{1}}\bigg{|}_{X=X_{\EW}} 
      &= 0 \implies \dfrac{\partial V}{\partial x_{1}}\bigg{|}_{X=X_{\EW}}  = -\dfrac{v_{2}^{2}}{2v_{1}v_{2}}\dfrac{\partial V}{\partial x_{3}}\bigg{|}_{X=X_{\EW}} \\
    \dfrac{\partial V}{\partial r_{2}}\bigg{|}_{X=X_{\EW}} 
      &= 0\implies \dfrac{\partial V}{\partial x_{2}}\bigg{|}_{X=X_{\EW}}  = -\dfrac{v_{1}^{2}}{2v_{1}v_{2}}\dfrac{\partial V}{\partial x_{3}}\bigg{|}_{X=X_{\EW}} \\
    \dfrac{\partial V}{\partial i_{1}}\bigg{|}_{X=X_{\EW}} &= 0 
      \implies \dfrac{\partial V}{\partial x_{4}}\bigg{|}_{X=X_{\EW}}  = 0\,.
\end{align}
We can therefore see that 
\begin{align}
    \nabla_{X}V\bigg{|}_{X=X_{\EW}} 
    = \left(-\dfrac{1}{2v_{1}v_{2}}\dfrac{\partial V}{\partial x_{3}}\bigg{|}_{X=X_{\EW}}\right)
    \begin{pmatrix}
        v_{2}^{2}\\
        v_{1}^{2}\\
        -2v_{1}v_{2}\\
        0
    \end{pmatrix}
\end{align}
and we note that this expression implies the following relation:
\begin{align}
    \left(X\cdot\nabla_{X}V\right)\bigg{|}_{X=X_{\EW}} = 0\,.
\end{align}
Combining this expression with \(\nabla_{X}V = A + B X\), we obtain 
\begin{align}
    X_{\EW}^{T}A + X_{\EW}^{T}BX_{\EW} = 0\,.
\end{align}
Therefore, the classical potential evaluated at the \(\EW\) extrema is equal to:
\begin{align}
    V_{\EW} = \dfrac{1}{2}A^{T}X_{\EW} = -\dfrac{1}{2}X_{\EW}^{T}BX_{\EW} 
\end{align}
In a similar manner, we find what the non-trivial minimisation conditions imply for \(X_{\CB}\):
\begin{align}
    \dfrac{\partial V}{\partial r_{1}}\bigg{|}_{X=X_{\CB}} 
      &= 0 \implies \dfrac{\partial V}{\partial x_{1}}\bigg{|}_{X=X_{\CB}}  = -\dfrac{\bar{v}_{2}^{2}}{2\bar{v}_{1}\bar{v}_{2}}\dfrac{\partial V}{\partial x_{3}}\bigg{|}_{X=X_{\CB}} \\
    \dfrac{\partial V}{\partial r_{2}}\bigg{|}_{X=X_{\CB}} 
      &= 0\implies \dfrac{\partial V}{\partial x_{2}}\bigg{|}_{X=X_{\CB}}  
      = -\dfrac{\bar{v}_{1}^{2}}{2\bar{v}_{1}\bar{v}_{2}}\dfrac{\partial V}{\partial x_{3}}\bigg{|}_{X=X_{\CB}} \\
    \dfrac{\partial V}{\partial c_{3}}\bigg{|}_{X=X_{\CB}} 
      &= 0\implies \dfrac{\partial V}{\partial x_{2}}\bigg{|}_{X=X_{\CB}}  
      = 0\\
    \dfrac{\partial V}{\partial i_{1}}\bigg{|}_{X=X_{\CB}} &= 0 
      \implies \dfrac{\partial V}{\partial x_{4}}\bigg{|}_{X=X_{\CB}}  = 0\,.
\end{align}
From these equations it is clear that 
\begin{align}
\nabla_{X}V\big{|}_{X=X_{\CB}} = 0 = A + BX_{\CB} \qquad \implies A = -BX_{\CB}
\end{align}
from which one obtains the value of the classical potential at
the \(\CB\) extrema, to wit
\begin{align}\label{eqn:VCB}
    V_{\CB} = \dfrac{1}{2}A^{T}X_{\CB} = -\dfrac{1}{2}X_{\CB}^{T}BX_{\CB}\,.
\end{align}
Combining the above results, we obtain:
\begin{align}\label{eqn:deriv_VCB}
    X_{\EW}^{T}BX_{\CB} = -X_{\EW}^{T}A = X_{\EW}^{T}BX_{\EW} = -2V_{\EW}
\end{align}
Using this result, Eqn.~(\ref{eqn:deriv_VCB}), Eqn.~(\ref{eqn:VCB}) and the fact that \(B = B^{T}\), we find that 
\begin{align}
    X_{\CB}^{T}\left(\nabla_{X}V\big{|}_{X=X_{\EW}}\right)
    &= X_{\CB}^{T}A + X_{\CB}^{T}BX_{\EW}\\ 
    &= -X_{\CB}^{T}BX_{\CB} + X_{\CB}^{T}BX_{\EW}\\
    &= 2V_{\CB} - 2V_{\EW}
\end{align}
We thus find that 
\begin{align}
    V_{\CB} - V_{\EW} = -\dfrac{1}{4v_{1}v_{2}}\left(\dfrac{\partial V}{\partial x_{3}}\bigg{|}_{X=X_{\EW}}\right)\left[(v_{1}\bar{v}_{1} - v_{2}\bar{v}_{2})^{2} + \alpha^{2}v_{1}^{2}\right]
\end{align}
Now suppose that \(V_{\EW}\) is a local minimum of the theory. It is possible to show that the mass of the charged Higgs is given by:
\begin{align}
    M^{2}_{H^{\pm}} = -\dfrac{v_{1}^{2} + v_{2}^{2}}{2v_{1}v_{2}}\dfrac{\partial V}{\partial x_{3}}\bigg{|}_{X=X_{\EW}}
\end{align}
Using this, we finally obtain that
\begin{align}
    V_{\CB} - V_{\EW} = \dfrac{M^{2}_{H^{\pm}}}{2(v_{1}^{2} + v_{2}^{2})}\left[(v_{1}\bar{v}_{1} - v_{2}\bar{v}_{2})^{2} + \alpha^{2}v_{1}^{2}\right]
\end{align}
The implications of this expression are clear: if the potential has coexisting EW and CB stationary points, and the EW solution is actually a minimum, then all of its squared scalar masses will necessarily be positive; therefore, since the quantity in square brackets is guaranteed to be positive, an EW minimum implies $V_{\CB} - V_{\EW}\,>\,0$ {\em and therefore the EW minimum is deeper than the CB stationary point}. Further, it can be shown that under these conditions the CB extremum is a saddle point. Thus, in refs.~\cite{Ferreira:2004yd,Barroso:2005sm,Maniatis:2006fs, Nishi:2006tg,Ivanov:2006yq,Ivanov:2007de,Maniatis:2007vn} the following tree-level theorem was established:
\begin{itemize}
    \item If the 2HDM tree-level scalar potential has an electroweak minimum, any charge breaking extremum that eventually exists will necessarily lie above that minimum.
    \item Further, the charge breaking extremum will necessarily be a saddle point.
\end{itemize}

We will now investigate these properties of the 2HDM vacuum structure at the loop level.

\section{One-Loop Corrections to the Scalar Potential}\label{sec:one_loop}

The classical scalar potential of a quantum field theory is not the true scalar potential. In an interacting quantum field theory, quantum effects will induce corrections to the scalar potential. The standard way of computing the corrections to the classical scalar potential is to use the path integral and background field method. For clarity, we will explain this formalism in a quantum field theory of a single interacting scalar field. We begin by writing down the so-called generating functional of the theory in terms of a path integral over field configurations:
\begin{align}
    Z[j] &= \mathcal{N}\int\mathcal{D}\Phi e^{\frac{i}{\hbar}S[\Phi] + \frac{i}{\hbar}\int d^{4}x j(x)\Phi(x)}
\end{align}
where $\Phi$ is the scalar field, $j$ is an external source and $S[\phi]$ is the action of the theory. By taking $n$-functional derivatives of $Z[j]$ with respect to $j$, one can generate the $n$-point Green's function consisting of the connected and disconnected Feynman diagrams with $n$-external propagators. We redefine the field $\Phi$ to consist of a classical component $\phi_{\cl}$ and a fluctuation field $\phi$: $\Phi = \phi_{\cl} + \hbar\phi(x)$. Here $\phi_{\cl}$ is chosen to satisfy the classical equations of motion in the presence of the external source $j$. We can expand the action of the theory around the classical field using:
\begin{align}
    S[\Phi] &= S[\phi_{\cl}] + \hbar\int d^{4}x \dfrac{\delta S}{\delta\Phi(x)}\phi(x) + \dfrac{\hbar^2}{2}\int d^{4}x\int d^{4}y \dfrac{\delta^2 S}{\delta\Phi(x)\delta\Phi(y)}\phi(x)\phi(y) + \cdots
\end{align}
where the $\cdots$ represent higher-order functional derivatives of the action (which aren't of interest to use here.) Using the expansion of the action in $Z[j]$, 
\begin{align}
    Z[j] &= \mathcal{N}e^{\frac{i}{\hbar}S[\phi_{cl}] + \frac{i}{\hbar}\int d^{4}x j(x)\phi_{\cl}}\int\mathcal{D}\Phi \exp\bigg{[}\dfrac{i\hbar}{2}\int d^{4}x\int d^{4}y \dfrac{\delta^2 S}{\delta\Phi(x)\delta\Phi(y)}\bigg{|}_{\phi_{cl}}\phi(x)\phi(y)\\
    &\hspace{6.5cm} + i\int d^{4}x \left(\dfrac{\delta S}{\delta\Phi(x)}\bigg{|}_{\phi_{cl}} + j\right)\phi(x) + \cdots\bigg{]} \notag
\end{align}
where again, the $\cdots$ represent higher-order terms. Since $\phi_{\cl}$ satisfies the classical equations of motion in the presence of the source $j$, the term linear in $\phi$ vanishes. The quadratic term can be integrated exactly, yielding:
\begin{align}
    Z[j] &= \mathcal{N}e^{\frac{i}{\hbar}S[\phi_{cl}] + \frac{i}{\hbar}\int d^{4}x j(x)\phi_{\cl}}\left[\det\dfrac{\delta^2 S}{\delta\Phi(x)\delta\Phi(y)}\right]^{-1/2}\left(1 + \order{\hbar}\right)
\end{align}
Next, we define $W[j] = -i\hbar \log(Z[j])$, which is the generating functional for the connected Green's functions. To order $\hbar$, this is:
\begin{align}\label{eqn:gen_func_w}
    W[j] &= -i\hbar\log(\mathcal{N}) + S[\phi_{\cl}] + \int d^{4}x j(x)\phi_{\cl} + \dfrac{i\hbar}{2}\log\left(\det\dfrac{\delta^2 S}{\delta\Phi(x)\delta\Phi(y)}\right) + \order{\hbar^2}
\end{align}
From now on, we will drop the $\log(\mathcal{N})$ term, since it a constant and will not play any role. Lastly, we define the effective action $\Gamma[\bar{\phi}]$ through the Legendre transform of $W[j]$:
\begin{align}
    \Gamma[\bar{\phi}] = W[j] - \int d^{4}x j(x)\bar{\phi}
\end{align}
where $\bar{\phi}(x) = \delta W[j]/\delta j(x)$. To order $\hbar$, the field $\bar{\phi}(x)$ is given by (using Eqn.~(\ref{eqn:gen_func_w})):
\begin{align}
    \dfrac{\delta W[j]}{\delta j} &= \dfrac{\delta S}{\delta\phi_{\cl}}\dfrac{\delta \phi_{\cl}}{\delta j} + \phi_{\cl} + j\dfrac{\delta\phi_{\cl}}{\delta j} + \order{\hbar} = \left(\dfrac{\delta S}{\delta\phi_{\cl}} + j\right)\dfrac{\delta \phi_{\cl}}{\delta j} + \phi_{\cl} + \order{\hbar} = \phi_{\cl} + \order{\hbar}
\end{align}
Using this relationship, we can write $\phi_{\cl} = \bar{\phi} + \phi_{1}$, where $\phi_{1}$ is of order $\hbar$. We can now replace $\phi_{\cl}$ in favor of $\bar{\phi}$. Given that $\delta S/\delta\phi\big{|}_{\phi=\phi_{\cl}} = -j$, we can think of $j$ as a functional of $\phi_{\cl}$, replacing $\phi_{\cl}$ in favor of $\bar{\phi}$. Writing $\Gamma_{1}[\phi_{\cl}] =i\hbar/2\log\det(\delta^2/\delta\Phi(x)\delta\Phi(y)\big{|}_{\phi=\phi_{\cl}})$, we find, to order $\hbar$:
\begin{align}
    \Gamma[\bar{\phi}] &= 
    S[\bar{\phi}]
    + \int d^4 x \phi_{1}(x)\dfrac{\delta S[\bar{\phi}]}{\delta\bar{\phi}(x)} 
    + \int d^4x(\bar{\phi} + \phi_{1})\left(j[\bar{\phi}] 
    + \phi_{1}\dfrac{\delta j}{\delta\bar{\phi}}\right) 
    + \Gamma_1[\bar{\phi}]\\
    &\qquad
    - \int d^4x \bar{\phi}\left(j[\bar{\phi}] + \phi_{1}\dfrac{\delta j}{\delta\bar{\phi}}\right) + \order{\hbar^2}\notag\\
    & = S[\bar{\phi}] + \Gamma_{1}[\bar{\phi}] + \order{\hbar}
\end{align}
where we dropped terms that go like $\order{\phi_{1}^2}$ since they are $\order{\hbar^2}$. If we take $\bar{\phi}$ to be space-time independent, the classical action evaluated at $\bar{\phi}$ is simply $-(VT)V_{0}(\bar{\phi})$ where $VT$ is the space-time volume and $V_{0}$ is the tree-level scalar potential. We thus define the effective potential as:
\begin{align}
    V_{\eff}(\bar{\phi}) &= -\dfrac{\Gamma[\bar{\phi}]}{VT} = V_{0}(\bar{\phi}) - \dfrac{i\hbar}{2(VT)}\log\left(\det\dfrac{\delta^2 S}{\delta\Phi(x)\delta\Phi(y)}\bigg{|}_{\phi=\bar{\phi}}\right) + \order{\hbar^2}
\end{align}
It is straightforward to evaluate the $\log(\det(\delta^2S/\delta\Phi\delta\Phi))$ term by using the identity $\log(\det(A)) = \tr(\log(A))$. For a real scalar field, one has $\delta^2S/\delta\Phi\delta\Phi = \Box + m^2(\bar{\phi})$ (where $m^2(\bar{\phi})$ is the field-dependent mass computed by diagonalizing $\partial^2V_{0}(\bar{\phi})/\partial\bar{\phi}^2$) and hence:
\begin{align}
    \tr\log(\Box + m^2) = VT\int\dfrac{d^{4}p}{(2\pi)^4}\log(-p^2+m^2)
\end{align}
The integral can be computed by replacing $\log(-p^2 + m^2)$ with $-\lim_{\alpha\to0}\frac{\partial}{\partial\alpha}(-p^2+m^2)^{-\alpha}$ and using standard one-loop integral tables. The result is divergent and requires the couplings of the theory to be renormalized. Once the infinities are canceled off (using $\overline{\mathrm{MS}}$), the result is:
\begin{align}
    - \dfrac{i\hbar}{2(VT)}\log\left(\det\dfrac{\delta^2 S}{\delta\Phi(x)\delta\Phi(y)}\bigg{|}_{\Phi=\bar{\phi}}\right) &= \dfrac{\hbar}{64\pi^2}m^4(\bar{\phi})\left[\log\left(\dfrac{m^2(\bar{\phi})}{\mu^2}\right)-\dfrac{3}{2}\right]
\end{align}
with $\mu$ being the renormalization scale. It is straight forward to add in additional scalar fields, gauge bosons, and fermions. The form of effective potential to $\order{\hbar}$ is~\cite{martin2002two}:
\begin{align}\label{eqn:hbar_exp_pot}
    V_{\eff}(\bar{\phi}) = V_{0}(\bar{\phi}) +  \dfrac{\hbar}{64\pi^{2}}\sum_{i}(-1)^{2s_{i}}n_{i}\left[M_{i}^{2}(\bar{\phi})\right]^{2}\left[\log(\dfrac{M_{i}^{2}(\bar{\phi})}{\mu^{2}}) - c_{i}\right] + \order{\hbar^2}
\end{align}
where $i$ runs over all the particles of the theory, \(s_{i}\) is the spin of the particle, \(n_{i}\) is the number of degrees of freedom of the particle, \(\mu\) is the renormalization scale and \(M_{i}^{2}(\bar{\phi})\) is the field dependent squared mass. The value of \(c_{i}\) is renormalization-scheme-dependent. For \(\overline{\mathrm{MS}}\)~\cite{martin2002two}, \(c_{i} = 5/6\) for gauge fields and \(3/2\) for all other particles. In principle, one needs to take into account the order $\hbar$ correction present in $\bar{\phi}$ when computing $V_{0}(\bar{\phi})$. The terms of order $\hbar$ arising from $V_{0}(\bar{\phi})$ play an important role in ensuring that the effective potential is gauge-independent order-by-order in $\hbar$. We will discuss this further in Sec.~(\ref{sec:hbar_expansion}). In the remaining subsections, we provide results for the various contributions from the particles involved in the 2HDM and discuss the $\hbar$ expansion of the effective potential.

\subsection{Scalar Contributions}

For the scalar fields, the one-loop correction to the scalar potential is
\begin{align}
    V^{(1)}(\Phi) = \dfrac{1}{64\pi^{2}}\sum_{i}\left[M_{i}^{2}(\Phi)\right]^{2}\left[\log(\dfrac{M_{i}^{2}(\Phi)}{\mu^{2}}) - \dfrac{3}{2}\right].
\end{align}
In the expression above, the values of the squared masses are the eigenvalues of the   second derivative of the tree-level potential, \(M^{2}_{ij}(\Phi)\):
\begin{align}
    M^{2}_{ij}(\Phi) = \frac{1}{2}\,\dfrac{\partial^{2}V^{(0)}}{\partial\phi_{i}\partial\phi_{j}}\left(\Phi\right)\,,
\end{align}
with $\{\phi_{i}, \phi_{j}\}$ any of the real components defined in eq.~\eqref{eqn:phi1_and_phi2}. In a general gauge, there are additional gauge dependent pieces which contribute to the scalar squared mass matrix. These gauge contribution have the effect of giving the Goldstones masses which are $\xi$ times the corresponding massive gauge bosons (see below for the gauge masses), where $\xi$ is the gauge-fixing parameter. However, we will chose the Landau gauge $\xi=0$, where the additional gauge-dependent pieces do not contribute to the scalar mass matrix. For a general field configuration, this \(8\times8\) mass matrix is extremely complicated, preventing us from giving explicit expressions to its eigenvalues. It is, however, possible to compute the scalar masses and their derivatives (which we will need) for the cases where $c_{2}=c_{3}=c_{4}=i_{1}=i_{2}=0$. These expressions are lengthy and we will, therefore, omit the results, having in any way developed a numerical procedure to obtain their values for our calculation.

\subsection{Gauge Contributions}

The field-dependent squared masses of the W and Z bosons and the photon are generated from the kinetic terms of the two Higgs doublets: $(D_{\mu}\Phi_{i})^\dagger(D^{\mu}\Phi_{i})$. Plugging in the expectation values for the Higgs doublets, the result is:
\begin{align}
    \mathcal{L}_{\mathrm{gauge,mass}} 
    &= 
    \dfrac{1}{4}g'^{2}B_{\mu}B^{\mu}\left(|\expval{\Phi_{1}}|^{2} + |\expval{\Phi_{2}}|^{2}\right) + g^{2}W_{\mu}^{a}W^{\mu,a}\left(|\expval{\Phi_{1}}|^{2} + |\expval{\Phi_{2}}|^{2}\right)\notag\\
    &\qquad + \dfrac{1}{2}gg'B_{\mu}W^{\mu,a}\left(\expval{\Phi_{1}}^{\dagger}\sigma^{a}\expval{\Phi_{1}} + \expval{\Phi_{2}}^{\dagger}\sigma^{a}\expval{\Phi_{2}}\right)
\end{align}
where \(g'\) and \(g\) are the \(U(1)_{\mathrm{Y}}\) and \(SU(2)_{\mathrm{L}}\) gauge couplings and $\sigma^{a} = \sigma^{1},\sigma^{2},\sigma^{3}$ are the Pauli-sigma matrices. In order to compute the squared masses of the gauge fields, it is useful to organize the gauge fields into the following vector:
\begin{align}
    \vec{G}_{\mu} = 
    \begin{pmatrix}
        W_{\mu}^{1} &
        W_{\mu}^{2} &
        W_{\mu}^{3} &
        B_{\mu}\\
    \end{pmatrix}^{T}
\end{align}
Computing the second derivative of $\mathcal{L}_{\mathrm{gauge,mass}}$ with respect to the components of $\vec{G}_{\mu}$ generates the following $4\times4$ mass squared matrix for the gauge bosons:
\begin{align}
    \dfrac{\partial^{2}\mathcal{L}_{\mathrm{gauge,mass}}}{\partial G^{i}_{\mu}\partial G^{j}_{\mu}} = M_{i,j}^{2}(\Phi) &= 
    \dfrac{g^2}{4}
    \begin{pmatrix}
        x + y & 0 & 0 & 2t_{W}z\\
        0 & x + y & 0 & 2t_{W}w\\
        0 & 0 & x + y & t_{W}(y-x)\\
        2t_{W}z & 2t_{W}w & t_{W}(y-x) & t_{W}^{2}(x + y)
    \end{pmatrix}
    \label{eq:matg}
\end{align}
where we have defined: \(t_{W} \equiv \tan(\theta_{W}) = g'/g\) and the parameters $x,y,z$ and $w$ as:
\begin{align}
    x &= r_{1}^{2} + r_{2}^{2} + i_{1}^{2} + i_{2}^{2}\\
    y &= c_{1}^{2} + c_{2}^{2} + c_{3}^{2} + c_{4}^{2}\\
    z &= c_{2}i_{1} + c_{4} i_{2} + c_{1} r_{1} + c_{3} r_{2}\\
    w &= c_{1}i_{1} + c_{3} i_{2} - c_{2} r_{1} - c_{4} r_{2}
\end{align}
It is possible to explicitly compute the eigenvalues of the gauge mass squared matrix. In terms of the above parameters, the squared masses of the $W, Z$ and photon are~\footnote{For charge preserving minima alone; otherwise, none of these four masses would be zero and there would not be states identified as charged or neutral since charge conservation would be broken.}:
\begin{align}
    M_{W}^{2} &= \dfrac{g^{2}}{4}(x+y)\\
    M_{Z}^{2} &= \frac{1}{8} g^2 \left(1 + t_{W}^2\right) \left[x+y+\sqrt{(x+y)^2 + 16 s_W^2c_W^2 \left(w^2-x y+z^2\right)}\right]\\
    M_{\gamma}^{2} &= \frac{1}{8} g^2 \left(1 + t_{W}^2\right) \left[x+y-\sqrt{(x+y)^2 + 16 s_W^2c_W^2 \left(w^2-x y+z^2\right)}\right]
\end{align}
with \(s_{W} \equiv \sin(\theta_{W})\) and \(c_{W} \equiv \cos(\theta_{W})\). In general, one also needs to consider the effects of ghosts. Ghost fields add additional contributions to the effective potential with squared mass equal to $\xi_{i} m^2_{g,i}$, for each of the massive gauge bosons. However, we will work in the $\xi = 0$ Landau gauge where the ghosts and Golstone bosons are massless.

\subsection{Top quark Contribution}

For simplicity, the only fermion we consider is the top quark. The contributions to the effective potential from other fermions will be significantly smaller than the top quark contribution since the top mass is almost two orders of magnitude greater than the next heaviest fermion. To compute the field-dependent squared mass of the top quark, we consider the Yukawa interactions between the Higgs doublets and the top quark:
\begin{align}
    \mathcal{L}_{\mathrm{Yukawa}} &= -y_{t}\bar{Q}_{L}\tilde{\Phi}t_{R} + \mathrm{h.c.} + \cdots
\end{align}
where \(\Phi = \Phi_{1}\) or \(\Phi_{2}\) and the \(\cdots\) terms represents Yukawa interactions involving the remaining fermions, which we ignore. Following the usual convention, we take $\Phi = \Phi_{2}$, for which the top quark mass is given by:
\begin{align}
    m_{t}^{2} &= \dfrac{1}{2}(c_{3}^2 + c_{4}^2+r_{2}^2+i_{2}^2)y_{t}^{2}
\end{align}
The Yukawa coupling $y_{t}$ will depend on the values we choose for the $r_{2}, i_{2}, c_{3}$ and $c_{4}$. We define the Yukawa coupling through the $\EW$ VEV, i.e. with $r_{2}=v_{2}$ and $i_{2} = c_{3} = c_{4} = 0$. The resulting Yukawa coupling is given by:
\begin{align}
    y_{t} = \dfrac{\sqrt{2}m_{t}}{|v_{2}|}
\end{align}
Given this Yukawa coupling, the field dependent top quark mass is:
\begin{align}
    m_{t}^{2} &= \dfrac{m_{t}^2}{v_{2}^2}(c_{3}^2 + c_{4}^2+r_{2}^2+i_{2}^2)
\end{align}
Given that we will not consider the (much smaller than the top's) contributions from other quarks or leptons, the results we present here will (within that approximation) therefore be valid for the several Yukawa-types of 2HDM (Type I, II, lepton-specific and flipped~\cite{Branco:2011iw}).

\subsection{\boldmath \(\hbar\)-Expansion}\label{sec:hbar_expansion}

As is well known, the effective potential is a gauge-dependent quantity. In principle, however, physical quantities calculated from the effective potential should {\em not} be gauge dependent. In practice, however, {\em how} such physical quantities are calculated determines whether or not the gauge dependence appears.

The theoretical backbone for these issues are the so-called Nielsen identities \cite{Nielsen:1975fs}, which can be cast as the fact that variations of the effective potential with respect to the gauge parameter $\xi$ are proportional to variations with respect to the field itself,
\begin{equation}
    \frac{\partial}{\partial \xi}V_{\rm eff}(\phi,\xi)=C(\phi,\xi)\frac{\partial}{\partial\phi}V_{\rm eff}(\phi,\xi).
\end{equation}
The equation above holds order by order in perturbation theory and, in particular, it implies that the value of $V_{\rm eff}$ at critical points, i.e. where $\partial V_{\rm eff}/\partial\phi=0$, is gauge independent.

The key issue with the ``brute force'' minimization of the effective potential to compute physical quantities lies with the fact that truncating the perturbative expansion means that incomplete higher-order terms are, implicitly, introducing a spurious gauge dependence. The proposal of Ref.~\cite{Patel:2011th}, known as the $\hbar$ expansion method, consists of casting the effective potential (and its derivatives) as a series in $\hbar$, after ``reintroducing'' the $\hbar$ in the partition function. The minimization is then carried out by an ``inversion of series'' method \cite{Patel:2011th}. Notice that while the $\hbar$-expansion method was originally developed for the finite-temperature effective potential, its applicability extends (in fact, in a much more straightforward way) to the zero-temperature effective potential we are concerned with here. 

The $\hbar$-expansion method is manifestly gauge-independent, and unlike ``brute force'' minimization, it does not introduce an imaginary part in the broken phase. Also, it is valid at all types of extrema, including maxima and saddle points. In practice, the method's prescription is simply to find the extrema of the tree-level potential, with the perturbative series generating the corrections order by order.

As mentioned in Eqn.~(\ref{eqn:hbar_exp_pot}), the effective potential can be expanded in terms of \(\hbar\). To be consistent, we must also include the order $\hbar$ contributions present in the vacuum configuration:
\begin{align}
    \vec{\phi}_{\mathrm{vac}} &= \vec{\phi}^{(0)}_{\mathrm{vac}} + \hbar\vec{\phi}^{(1)}_{\mathrm{vac}} + \mathcal{O}(\hbar^{2})
\end{align}
The effects of including the $\order{\hbar}$ contribution to the vacuum configuration is to introduce additional terms arising from the tree-level potential that contribute to the effective potential at order $\hbar$. The full scalar potential evaluated at \(\vec{\phi}_{\mathrm{vac}}\), expanded to order \(\mathcal{O}(\hbar^{2})\) is:
\begin{align}
    V_{\mathrm{eff}}(\vec{\phi}_{\mathrm{vac}}) &= V^{(0)}(\vec{\phi}^{(0)}_{\mathrm{vac}}) + \hbar \left[V^{(1)}(\vec{\phi}^{(0)}_{\mathrm{vac}}) + \sum_{k}\vec{\phi}^{(1)}_{\mathrm{vac},k}\dfrac{\partial V^{(0)}}{\partial\phi_{k}}(\vec{\phi}^{(0)}_{\mathrm{vac}})\right]
\end{align}
The extrema conditions for the full effective potential are then given by:
\begin{align}\label{eqn:veff_extrema}
    \dfrac{\partial V_{\mathrm{eff}}}{\partial\phi_{n}}(\vec{\phi}_{\mathrm{vac}}) &= \dfrac{\partial V^{(0)}}{\partial\phi_{n}}(\vec{\phi}^{(0)}_{\mathrm{vac}}) + \hbar \left[\dfrac{\partial V^{(1)}}{\partial\phi_{n}}(\vec{\phi}^{(0)}_{\mathrm{vac}}) + \sum_{k}\vec{\phi}^{(1)}_{\mathrm{vac},k}\dfrac{\partial^{2} V^{(0)}}{\partial\phi_{k}\partial\phi_{n}}(\vec{\phi}^{(0)}_{\mathrm{vac}})\right]
\end{align}
From this expression, we can immediately interpret the meaning of \(\vec{\phi}^{(0)}_{\mathrm{vac}}\): if \(\vec{\phi}_{\mathrm{vac}}\) is an extrema of \(V_{\mathrm{eff}}\), then \(\vec{\phi}^{(0)}_{\mathrm{vac}}\) is a vacuum configuration that extremizes the classical scalar potential.  Eqn.~(\ref{eqn:veff_extrema}) also shows us how to find the extrema of the full effective scalar potential to order \(\hbar\). One simply needs to determine all the extrema of the tree-level potential. Then, setting the term of order \(\hbar\) in Eqn.~(\ref{eqn:veff_extrema}) to zero, we can determine the one-loop correction to the classical vacuum configuration, \(\vec{\phi}^{(1)}_{\mathrm{vec}}\). Once the classical extrema have been determined, the minimum of the effective scalar potential will be the configuration which gives the smallest value of 
\begin{align}
    V_{\mathrm{eff}}(\vec{\phi}_{\mathrm{vac}}) &= V^{(0)}(\vec{\phi}^{(0)}_{\mathrm{vac}}) + \hbar V^{(1)}(\vec{\phi}^{(0)}_{\mathrm{vac}}) 
\end{align}
Note we have dropped \(\dfrac{\partial V^{(0)}}{\partial\phi_{k}}(\vec{\phi}^{(0)}_{\mathrm{vac}})\) since \(\vec{\phi}^{(0)}_{\mathrm{vac}}\) extremizes the classical scalar potential.

Ref.~\cite{Elias-Miro:2014pca} and \cite{Martin:2014bca} revealed IR divergences in the Landau gauge arising from massless Goldstone bosons, and argued that a resummation is necessary. However, e.g. Ref.~\cite{Ekstedt:2018ftj} argued that the $\hbar$-expansion obviates the need for a resummation because the IR divergences cancel order by order in perturbation theory. Notice that this procedure was extended in Ref.~\cite{Andreassen:2014eha} to the small mass limit. In the case of small, non-Goldstone masses, a resummation is necessary. If negative masses are found corresponding to a one-loop minimum, which would always be the case in our theory (because remember, a tree-level EW minimum implies that any coexisting tree-level CB extremum is a saddle point), one would additionally need to perform a resummation of the two-point correlation functions to obtain a more accurate result for the scalar masses. In the $\hbar$-expansion method, attempting to find a counterexample to the tree-level theorem, {\em e.g.} simultaneous  minima at one-loop, would always result in at least one set of negative squared masses (since either the $\EW$ or $\CB$ vacuum would be a saddle point at tree-level) and thus one would be left with imaginary one-loop potentials. Therefore, using the $\hbar$-expansion method to find counterexamples to the tree-level theorem would always require a resummation to provide a sensible result. We, therefore, will instead perform a numerical minimization of the effective potential and require that the tree-level potential is convex at one-loop minima, the one-loop effective potential thus becoming free of any imaginary pieces. Any gauge dependence of the results will be residual, stemming from the truncated perturbative expansion and arising, at least at the two-loop level, at order $\order{\hbar^2}$.

\section{Numerical Methods}\label{sec:num}

In this section, we describe the procedure we employ in finding counterexamples to the tree-level theorem on EW vacuum stability against charge breaking at one-loop order. A counterexample to the tree-level theorem is obtained if we can find a set of parameters for which there exist simultaneous \(\EW\) and \(\CB\)  minima - this, at tree-level, is impossible. Further, we will show that one-loop EW minima {\em may have deeper CB minima} and thus their stability is not guaranteed. In brief, the algorithm we use to find counterexamples is as follows:

\begin{enumerate}
    \item Generate $\EW$ and $\CB$ VEVs for $\Phi_{1}$ and $\Phi_{2}$ by sampling from 
    a uniform distribution.
    \item Generate initial random guesses for all eight of the 2HDM parameters: $m_{11}^{2}$,
    $m_{22}^{2}$, $m_{12}^{2}$, $\lambda_{1}$, $\lambda_{2}$, $\lambda_{3}$, $\lambda_{4}$ and
    $\lambda_{5}$ by sampling from uniform distributions. These will be used later as initial 
    ``seeds" for a numerical minimization of the potential.
    \item Extremize the effective potential at both the $\EW$ and $\CB$ by solving 
    the following five non-linear root equations:
    \begin{align}
        0=\dfrac{\partial V_{\eff}}{\partial r_{1}}\bigg{|}_{\phi_{\EW}} = 
    \dfrac{\partial V_{\eff}}{\partial r_{2}}\bigg{|}_{\phi_{\EW}} = 
    \dfrac{\partial V_{\eff}}{\partial r_{1}}\bigg{|}_{\phi_{\CB}} = 
    \dfrac{\partial V_{\eff}}{\partial r_{2}}\bigg{|}_{\phi_{\CB}} = 
    \dfrac{\partial V_{\eff}}{\partial c_{1}}\bigg{|}_{\phi_{\CB}}.
    \end{align}
    We solve these equations by holding the $\EW$ and $\CB$ VEVs fixed and varying five of the 2HDM parameters. We randomly chose which of the five 2HDM parameters we use to solve these equations each time.
    \item Choose a set of 50 random vacuua and perform minimizations at each to find remaining extrema.
    \item Categorize all of the extrema (as minimums, maximums or saddle points) by computing the eigenvalues of the effective potential Hessian. 
\end{enumerate}
Below, we describe the algorithm is more detail. Note that all the code was written in \mil{Julia} and is available on  \href{https://github.com/LoganAMorrison/THDMMinimizer.jl}{GitHub}.

\paragraph{\underline{\textbf{Randomly choosing VEVs:}}} Our starting point is to choose $\EW$ and $\CB$ vacuua at which we will attempt to extremize the effective potential. We characterize the mass scale of our problem in terms of the renormalization scale $\mu$ which we set to be the SM Higgs VEV: $\mu = 246~\mathrm{GeV}$. Since the effective potential contains logarithms of the form $\log(M^2/\mu^2)$, we choose all of our dimensionful parameters to be of the order of the renormalization scale. We do this to avoid unwanted large logarithms which can ultimately spoil our perturbative expansion. As we did in Sec.~(\ref{sec:tree}), we define the $\EW$ and $\CB$ vacuua as:
\begin{align}
    \expval{\Phi_{1}}_{\EW} &= 
    \dfrac{1}{\sqrt{2}}\begin{pmatrix}
        0\\
        v_{1}
    \end{pmatrix}, & 
    \expval{\Phi_{2}}_{\EW} &= 
    \dfrac{1}{\sqrt{2}}\begin{pmatrix}
        0\\
        v_{2}
    \end{pmatrix},\\
    \expval{\Phi_{1}}_{\CB} &= 
    \dfrac{1}{\sqrt{2}}\begin{pmatrix}
        \alpha\\
        \bar{v}_{1}
    \end{pmatrix}, & 
    \expval{\Phi_{2}}_{\CB} &= 
    \dfrac{1}{\sqrt{2}}\begin{pmatrix}
        0\\
        \bar{v}_{2}
    \end{pmatrix}.
\end{align}
In terms of the individual components of the fields $\Phi_{1}$ and $\Phi_{2}$ (see Eqn.~(\ref{eqn:phi1_and_phi2})), this means that the following real components
will have non-zero VEVs,
\begin{align}
    \expval{r_{1}}_{\EW} &= v_{1}, & \expval{r_{2}}_{\EW} &= v_{2}, &&\\
    \expval{r_{1}}_{\CB} &= \bar{v}_{1}, & \expval{r_{2}}_{\CB} &= \bar{v}_{2}, & \expval{c_{1}}_{\CB} &= \alpha
\end{align}
with all other component fields of Eqn.~(\ref{eqn:phi1_and_phi2})) have expectation values equal to zero. As stated above, we choose the scale of the VEVs to be on the order of the renormalization scale. That is, we set:
\begin{align}
    v_{1}^2 + v_{2}^2 &= (246\,\mbox{GeV})^2, & -\mu \leq \bar{v}_{1}, \bar{v}_{2}, \alpha &\leq \mu
\end{align}
We set $v_{1}^2 + v_{2}^2 = \mu^2 = (246\,\mbox{GeV})^2$ to of course in order obtain a SM-like $\EW$ vacuum, with gauge boson and quark masses in accordance wit experiment, but with an arbitrary value of $\tan\beta = v_{2}/v_{1}$. However, when we search for other minima by numerically minimizing the effective potential w.r.t. the fields $r_{2},r_{2}$ and $c_{1}$ (see below), we may find deeper $\EW$ minima which no longer satisfy this condition $v_{1}^2 + v_{2}^2 = (246\,\mbox{GeV})^2$ (this is a well known property of the 2HDM, already occurring at tree level). However, this condition allows us to find situations where at least there is a SM-like vacuum.

\paragraph{\underline{\textbf{Initializing the 2HDM Parameters:}}} In the 2HDM we consider, there are a total of 3 dimensionful mass parameters and 5 dimensionless quartic couplings (see Eqn.~(\ref{eqn:vtree})): 
\begin{align}
     m_{11}^2, m_{22}^2, m_{12}^2, \lambda_{1}, \lambda_{2}, \lambda_{3}, \lambda_{4}, \lambda_{5}
\end{align}
As with the vacuua, we choose the dimensionful mass parameters to be of the same order as the renormalization scale. That is, we choose:
\begin{align}
    -\mu^2 \leq &m_{11}^{2}, m_{22}^{2}, m_{12}^{2} \leq \mu^2.
\end{align}
As we stated above, we make this choice to avoid generating large scalar masses which in turn could lead to large logarithms. In choosing the values of the dimensionless couplings, we keep in mind that sufficiently large couplings will result in a breakdown of perturbation theory. In practice, the breakdown occurs when dimensionless expansion parameters exceed $4\pi$ (since the perturbative expansion is in powers of $(\text{expansion parameter})/4\pi$.) To satisfy perturbative unitarity, we keep all of the quartic couplings to be below 10. In addition to perturbative unitarity, we also wish to have a stable potential for the scalars. The tree-level conditions for stability of the scalar potential are:
\begin{align}
    \label{eqn:tree_bounded1}
    0\leq \lambda_{1},\lambda_{2},\\
    \label{eqn:tree_bounded2}
    -\sqrt{\lambda_{1}\lambda_{2}} \leq \lambda_{3},\\
    \label{eqn:tree_bounded3}
    -\sqrt{\lambda_{1}\lambda_{2}} \leq \lambda_{3} + \lambda_{4} - |\lambda_{5}|
\end{align}
With these conditions and perturbative unitarity in mind, we choose the quartic couplings such that:
\begin{align}
    0 \leq &\lambda_{1}, \lambda_{2} \leq 10,\\
    -\sqrt{\lambda_{1}\lambda_{2}} \leq &\lambda_{3} \leq 10 + \sqrt{\lambda_{1}\lambda_{2}}\\
    -1 \leq &\lambda_{4}, \lambda_{5} \leq 1 .
\end{align}
Even with these choices, it is possible to violate the stability conditions. Thus, we generate parameters according to the above prescriptions and then check if the three-level potential is bounded. If it is, we continue, otherwise, we continue to generate parameters until the potential is stabilized.

\paragraph{\underline{\textbf{Extremize the Effective Potential:}}} Our goal is ultimately to have minima at the $\EW$ and $\CB$ vacuua we have chosen. As a first step, we simultaneously extremize (not knowing ahead of time whether or not we are at a minimum, maximum or saddle point) the effective potential at the $\EW$ and $\CB$ vacuua. To do this, we must simultaneously solve the following five root equations:
\begin{align}\label{eqn:veff_extremal}
    0=\dfrac{\partial V_{\eff}}{\partial r_{1}}\bigg{|}_{\phi_{\EW}} = 
    \dfrac{\partial V_{\eff}}{\partial r_{2}}\bigg{|}_{\phi_{\EW}} = 
    \dfrac{\partial V_{\eff}}{\partial r_{1}}\bigg{|}_{\phi_{\CB}} = 
    \dfrac{\partial V_{\eff}}{\partial r_{2}}\bigg{|}_{\phi_{\CB}} = 
    \dfrac{\partial V_{\eff}}{\partial c_{1}}\bigg{|}_{\phi_{\CB}}.
\end{align}
The derivatives of the effective potential are given by:
\begin{align}\label{eqn:veff_deriv}
    \dfrac{\partial V_{\eff}}{\partial\phi}(\Phi) &= \dfrac{\partial V_{\tree}}{\partial\phi}(\Phi) + \dfrac{1}{32\pi^2}\sum_{i}\dfrac{\partial M^{2}_{s,i}(\Phi)}{\partial\phi}M^{2}_{s,i}(\Phi)\left[\log\left(\dfrac{M^{2}_{s,i}(\Phi)}{\mu^2}\right) - 1\right]\\
    &\qquad + \dfrac{1}{32\pi^2}\sum_{i}\dfrac{\partial M^{2}_{g,i}(\Phi)}{\partial\phi}M^{2}_{g,i}(\Phi)\left[3\log\left(\dfrac{M^{2}_{g,i}(\Phi)}{\mu^2}\right) - 1\right]\notag\\
    &\qquad - \dfrac{12}{32\pi^2}\dfrac{\partial M^{2}_{\mathrm{top}}(\Phi)}{\partial\phi}M^{2}_{\mathrm{top}}(\Phi)\left[\log\left(\dfrac{M^{2}_{\mathrm{top}}(\Phi)}{\mu^2}\right) - 1\right]\notag
\end{align}
Here $M^{2}_{s,i}(\Phi)$ are the eigenvalues of the scalar squared-mass matrix, $M^{2}_{g,i}(\Phi)$ are the eigenvalues of the gauge squared-mass matrix and $M^{2}_{\mathrm{top}}(\Phi)$ is the squared top mass. Note that the factor of $3$ on the $\log$ of the gauge contribution comes from the polarization of the massive gauge fields (the fact that the W boson is charged is also taken into account on the sum over the four eigenvalues of the gauge boson mass matrix of Eqn~\eqref{eq:matg}) and the factor of 12 for the top contribution accounts for the 3 colors, 2 spins, and charge of that particle.

To solve the five root equations, we must have five independent parameters which can vary. Since we wish to fix the $\EW$ and $\CB$ vacuua, we must resort to varying five of the 2HDM parameters. To ensure that we can sample the entire parameter space, we randomly choose any five 2HDM parameters ({\em i.e.}, any given five of the quadratic or quartic parameters) to vary each time we solve the extremal equations. We employ the \mil{NLsolve.jl} \mil{Julia} library \cite{nlsolve} using the Trust Region method. Since we allow five of the 2HDM parameters to vary, we could potentially find solutions which make the scalar potential unstable or spoil the perturbative expansion. We thus reject solutions which for which the stability conditions are violated or solutions which have 2HDM parameter which are too large ($m_{ij}^2 > (10\mu)^2$ or $|\lambda_{i}| > 10.$)

As explained in Sec.~(\ref{sec:one_loop}), there are no analytical expressions for the squared scalar masses for an arbitrary vacuum configuration. They must, therefore, be computed numerically by calculating the eigenvalues of the scalar squared mass matrix. This makes computing the derivatives of the eigenvalues of the scalar mass matrix extremely difficult. To obtain those (first and second-order) derivatives, then, we employ an algorithm using forward-mode automatic differentiation through the use of dual-numbers, which we explain in App.~(\ref{app:autodiff}). We use the \mil{FowardDiff.jl} package~\cite{RevelsLubinPapamarkou2016}, which implements a dual-number type in \mil{Julia}. This allows us to simply pass dual-number types into the effective potential, and we obtain automatic derivatives without ever needing to use Eqn.~(\ref{eqn:veff_deriv})\footnote{Currently, \mil{Julia} implements its linear algebra by calling LAPACK which doesn't accept any types other than floating-point numbers. Thus, we wrote a version of the Jacobi algorithm for computing eigenvalues.}.

\paragraph{\underline{\textbf{Finding Additional Minima:}}} As explained above, we solve the minimization conditions of the one-loop effective potential so as two obtain two different extrema. But beyond those two vacuua - one EW breaking, the other CB - the 2HDM potential may yet have other extrema. For instance, if at tree-level the CB minimum is unique (see~\cite{Ferreira:2004yd,Barroso:2005sm}) other neutral minima may exist (~\cite{Barroso:2007rr,Ivanov:2006yq,Ivanov:2007de,Branchina:2018qlf}). To search for any remaining minima of the effective potential, we randomly generate 50 vacuum configurations and perform a numerical minimization starting from these vacuua, verifying whether the potential may assume deeper values than the starting points. The minimization is performed using the Broyden–Fletcher–Goldfarb–Shanno algorithm provided from the \mil{Optim.jl} library~\cite{mogensen2018optim}. After performing this procedure, we sometimes find a minimum which is not one of the initial solutions found by solving the extremal equations of Eqn.~(\ref{eqn:veff_extremal}). For these deeper neutral minima, it is likely that we will now have $v_{1}^2 + v_{2}^2 \neq 246^2 \ \mathrm{GeV}^2$, the so-called ``panic vacuua" of refs.~\cite{Barroso:2012mj,Barroso:2013awa}.

\paragraph{\underline{\textbf{Characterizing Extrema:}}} After we have found extrema of the effective potential, we need to determine if they are minima, maxima or saddle points. In general, an extremum of a scalar function can be characterized by computing the eigenvalues of the Hessian matrix. The Hessian matrix is the matrix consisting of all second derivatives of the function, which in our case is an $8\times8$ matrix with components: $\partial^2 V_{\eff}/\partial\phi_{i}\partial\phi_{j}$. The components of the Hessian matrix of the effective potential are given by:
\begin{align}\label{eqn:veff_hess}
    \dfrac{\partial^2 V_{\eff}}{\partial \phi_{i}\partial \phi_{j}}(\Phi) &= \dfrac{\partial^2 V_{\tree}}{\partial \phi_{i}\partial \phi_{j}}(\Phi) + \dfrac{1}{32\pi^2}\sum_{i}\left\{M^{2}_{s,i}(\Phi)\dfrac{\partial^2 M^{2}_{s,i}(\Phi)}{\partial\phi_{i}\partial\phi_{j}}\left[\log\left(\dfrac{M^{2}_{s,i}(\Phi)}{\mu^2}\right) - 1\right]\right.\\
    &\hspace{4.5cm} + \left. \dfrac{\partial M^{2}_{s,i}(\Phi)}{\partial\phi_{i}}\dfrac{\partial M^{2}_{s,i}(\Phi)}{\partial\phi_{j}}\log\left(\dfrac{M^{2}_{s,i}(\Phi)}{\mu^2}\right)
    \right\}\notag\\
    &\hspace{0.75cm} + \dfrac{1}{32\pi^2}\sum_{i}\left\{
    M^{2}_{g,i}(\Phi)\dfrac{\partial^2 M^{2}_{g,i}(\Phi)}{\partial\phi_{i}\partial\phi_{j}}\left[3\log\left(\dfrac{M^{2}_{g,i}(\Phi)}{\mu^2}\right) - 1\right]\right.\notag\\
    &\left. \hspace{3cm} + \dfrac{\partial M^{2}_{g,i}(\Phi)}{\partial\phi_{i}}\dfrac{\partial M^{2}_{g,i}(\Phi)}{\partial\phi_{j}}\left[3\log\left(\dfrac{M^{2}_{g,i}(\Phi)}{\mu^2}\right) + 2\right]\right\}\notag\\
    &\hspace{0.75cm} +  
    \dfrac{1}{32\pi^2}\left\{ M^{2}_{\mathrm{top}}(\Phi)\dfrac{\partial^2 M^{2}_{\mathrm{top}}(\Phi)}{\partial\phi_{i}\partial\phi_{j}}\left[\log\left(\dfrac{M^{2}_{\mathrm{top}}(\Phi)}{\mu^2}\right) - 1\right]\right.\notag\\
    &\hspace{3cm} + \left. \dfrac{\partial M^{2}_{\mathrm{top}}(\Phi)}{\partial\phi_{i}}\dfrac{\partial M^{2}_{\mathrm{top}}(\Phi)}{\partial\phi_{j}}\log\left(\dfrac{M^{2}_{\mathrm{top}}(\Phi)}{\mu^2}\right)\right\}\notag
\end{align}
If the Hessian matrix is positive semi-definite (i.e. all the eigenvalues are greater than or equal to zero), then the extremum is a minimum (note that the zero eigenvalues signal a flat direction, which we will explain momentarily.) Similarly, if the eigenvalues of the Hessian are negative semi-definite or neither positive nor negative semi-definite, then the extremum is a maximum or saddle point, respectively. When the two Higgs doublets attain their non-trivial VEVs, the $\mathrm{SU}(3)_{\mathrm{c}}\times\mathrm{SU}(2)_{\mathrm{L}}\times\mathrm{U}(1)_{\mathrm{Y}}$ gauge group is broken. In the case of the $\EW$ VEVs, the gauge group is broken down to $\mathrm{SU}(3)_{\mathrm{c}}\times\mathrm{U}(1)_{\mathrm{EM}}$ and in the case of $\CB$ VEVs, the gauge group is broken down to $\mathrm{SU}(3)_{\mathrm{c}}$. In either case, we expect there to be Goldstone bosons corresponding to each broken generator of the gauge-group. The Goldstone bosons will manifest themselves as zero eigenvalues\footnote{This is the case for the particular gauge choice of $\xi=0$. If $\xi\neq0$, then the Goldstone squared masses will be $\xi m^2_{g,i}$ where $m_{g,i}$ are the masses of the gauge fields.} of the Hessian matrix, i.e. flat directions of the effective potential. In Tab.~(\ref{tab:extrema_type}), we list the various extrema type corresponding to the eigenvalues of the effective potential Hessian. As with computing first derivatives of the effective potential, to compute the second derivatives, we use automatic differentiation. This again allows us to simply pass dual-numbers into the effective potential (in this case we pass nested dual numbers, i.e. dual numbers consisting of dual-numbers, see App.~(\ref{app:autodiff})) and we obtain the second derivatives without ever having to use Eqn.~(\ref{eqn:veff_hess}).
\begin{table}[ht]\centering
    \begin{tabular}{ll}
        \hline
        $\partial^2 V_{\eff}/\partial\phi_{i}\partial\phi_{j}$ Eigenvalues & Extrema Type\\
        \hline
        3 zero, 5 positive & $\EW$ minimum\\
        3 zero, 5 negative & $\EW$ maximum\\
        3 zero, 5 positive and negative & $\EW$ saddle\\
        4 zero, 4 positive & $\CB$ minimum\\
        4 zero, 4 negative & $\CB$ maximum\\
        4 zero, 4 positive and negative & $\CB$ saddle\\
        \hline
    \end{tabular}
    \caption{Characterization of the extrema of the 2HDM effective potential.}
    \label{tab:extrema_type}
\end{table}
Notice that the second derivatives of the one-loop effective potential provide the one-loop squared scalar masses computed at zero external momentum. For massive scalars they are therefore an approximation to the exact result, but for the massless Goldstones - which must be computed at precisely zero external momentum - they yield the exact result. Obtaining the correct number of massless Goldstones for either the EW or CB extrema is a powerful check of our calculations.

\section{Results}\label{sec:results}

In this section, we describe the results of running the algorithm described in the previous section to find counter-examples to the tree-level theorem described in Sec.~(\ref{sec:tree}). To wit, our purpose is to investigate whether at the one-loop level an EW minimum is guaranteed to be stable against charge breaking -- {\em i.e.}, whether still at one-loop there is no deeper CB extremum. Further, we will verify whether at one-loop the existence of an EW minimum also implies that any CB extremum must need be a saddle point. All computations were run on a 2015 Mac Book Pro using 8 threads. We developed all of the code for this algorithm using the \mil{Julia} language, using various well-developed \mil{Julia} packages. For example, we use the \mil{ForwardDiff.jl}~\cite{RevelsLubinPapamarkou2016} package for automatic differentiation, \mil{NLsolve.jl}~\cite{nlsolve} for solving the root equations of Eqn.~(\ref{eqn:veff_extrema}) and \mil{Optim.jl}~\cite{mogensen2018optim} for performing minimizations. All the code developed for this project can be viewed/downloaded on \href{https://github.com/LoganAMorrison/THDMMinimizer.jl}{GitHub}. For more details, the interested reader may e-mail the authors.

To consider a set of vacuua ($\EW$ and $\CB$) and 2HDM parameters to yield a counter-example to the tree-level theorem, we set various requirements. First, to be a counter-example to the tree-level theorem, we must have a minimum of the effective potential at an $\EW$ \textit{and} $\CB$ vacuum. We consider a vacuum to be an extremum of the effective potential if the infinity-norm of the gradient is less than $10^{-5}$ (although in many cases we obtain much higher accuracy.) We categorize the extremal type (minimum, maximum or saddle) of a vacuum using the conditions in Tab.~(\ref{tab:extrema_type}). In particular, we consider an $\EW$ vacuum a minimum if 
the Hessian of the effective potential evaluated at the vacuum contains three zero masses (Goldstone bosons corresponding to the breaking of $\mathrm{SU}(2)_L\times\mathrm{U}(1)_Y\to\mathrm{U}(1)_{\mathrm{EM}}$) and five positive masses. In the case of a $\CB$ vacuum, we require four zero masses (the additional zero mass due to the explicit breaking of $\mathrm{U}(1)_{\mathrm{EM}}$) and four positive masses. In addition, we require that the tree-level potential is bounded from below (following the conditions of Eqns.~(\ref{eqn:tree_bounded1}-\ref{eqn:tree_bounded2}).) We also require that the values of the 2HDM be constrained to be of natural order: $|m_{ij}^2| < (10\mu)^2, |\lambda_{i}| < 10$ where $\mu$ is the renormalization scale. We make these requirements to preserve our perturbative expansion and avoid generating large masses which could result in large logarithms.

After running our algorithms for roughly 24 hours, we found $\sim3000$ sets of parameters which yield simultaneous one-loop $\EW$ and $\CB$ minima -- {\em this is the first demonstration that the tree-level vacuum stability theorem is no longer valid at one-loop}. Out of these 3000 sets, for $\sim1000$ of then, the global minimum of the one-loop effective potential was the $\CB$ vacuum; the remaining $\sim2000$ had the $\EW$ vacuum as the global minimum. To get a sense of how common the sets of parameters yielding counter-examples were, we also recorded those sets of parameters for which there was only an $\EW$ minimum (no $\CB$) and for which there was only a $\CB$ minimum (no $\EW$). The former yielded $\sim54000$ sets of parameters, while the latter $\sim17000$. Thus, we can see that parameters which yield both a $\CB$ and an $\EW$ minimum are roughly $5\%$ of those which yield a single minimum -- thus even at one-loop, we can expect that the exclusion of regions of parameter space due to CB vacuum instability will be rare. Furthermore, only 4 out of the 1000 points which yielded a deeper $\CB$ minimum have positive tree-level masses and 10 for the case where the $\EW$ was deeper (the remaining contained at least one negative tree-level mass from either the $\CB$ or $\EW$ vacuum.) We should stress, however, that our purpose is not to perform a thorough scan of the 2HDM parameter space to find charge breaking bounds of the model, but rather to prove that the tree-level vacuum stability theorem no longer holds. As mentioned in Sec.~(\ref{sec:hbar_expansion}), for parameters with negative tree-level masses which result in one-loop minima, one likely needs to perform a resummation to obtain a sensible result (i.e. one that doesn't exhibit an apparent instability - imaginary part of the effective potential), which we have not done. But having performed the scan over the model's parameters such that the tree-level masses at the one-loop minima were always positive, the issue of a complex one-loop effective potential is no longer an issue that should worry us. 
\begin{table}[ht]\centering
    \begin{tabular}{lll}
        \hline
        & $\CB$ Deepest & $\EW$ Deepest\\
        \hline
        $m_{11}^2$ & -47729.7 & -56573.3\\
        $m_{12}^2$ & 2062.2 & -3666.6\\
        $m_{22}^2$ & -25134.0 & -41089.7\\
        \hline
        $\lambda_{1}$ & 3.3 & 2.8\\
        $\lambda_{2}$ & 0.8 & 1.4\\
        $\lambda_{3}$ & 1.7 & 2.0\\
        $\lambda_{4}$ & 1.5 & 0.9\\
        $\lambda_{5}$ & -0.02 & 0.5\\
        \hline
        $y_{t}$ & 1.00 & 1.00\\
        \hline
        $v_{1}$ & -11.4 & -20.8 \\
        $v_{2}$ & -245.7 & 245.1\\
        \hline
        $\bar{v}_{1}$ & 63.4   & -196.3\\
        $\bar{v}_{2}$ & 43.9   & 26.9\\
        $\alpha$      & -161.1 & -56.0\\
        \hline
        $V_{\eff}(\phi_{\EW})$  & $-3.68\times 10^8$ & $-6.17\times10^{8}$ \\
        $V_{\eff}(\phi_{\CB})$ & $-3.72\times 10^8$ & $-6.07\times10^{8}$\\
        \hline
    \end{tabular}
    \caption{Parameter values. Quadratic parameters in GeV$^2$, VEVs in GeV and
    potential values in GeV$^4$. All values have been rounded for readability.}
    \label{tab:param_vals}
\end{table}

\begin{figure}[ht]
    \centering
    \begin{subfigure}[b]{0.48\textwidth}
        \includegraphics[width=\textwidth]{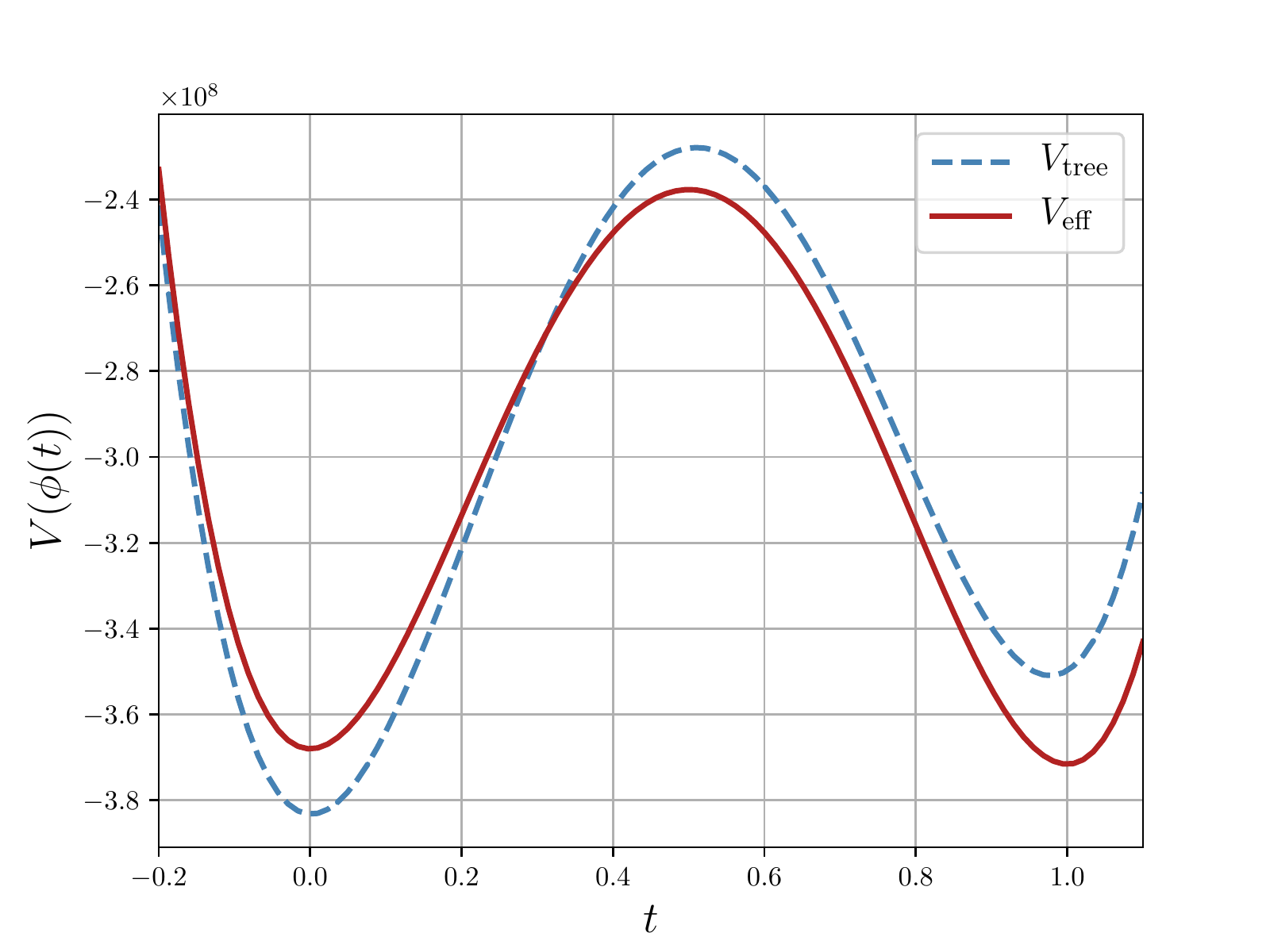}
        \caption{}
        \label{fig:pot_1D_cb}
    \end{subfigure}
    \begin{subfigure}[b]{0.48\textwidth}
        \includegraphics[width=\textwidth]{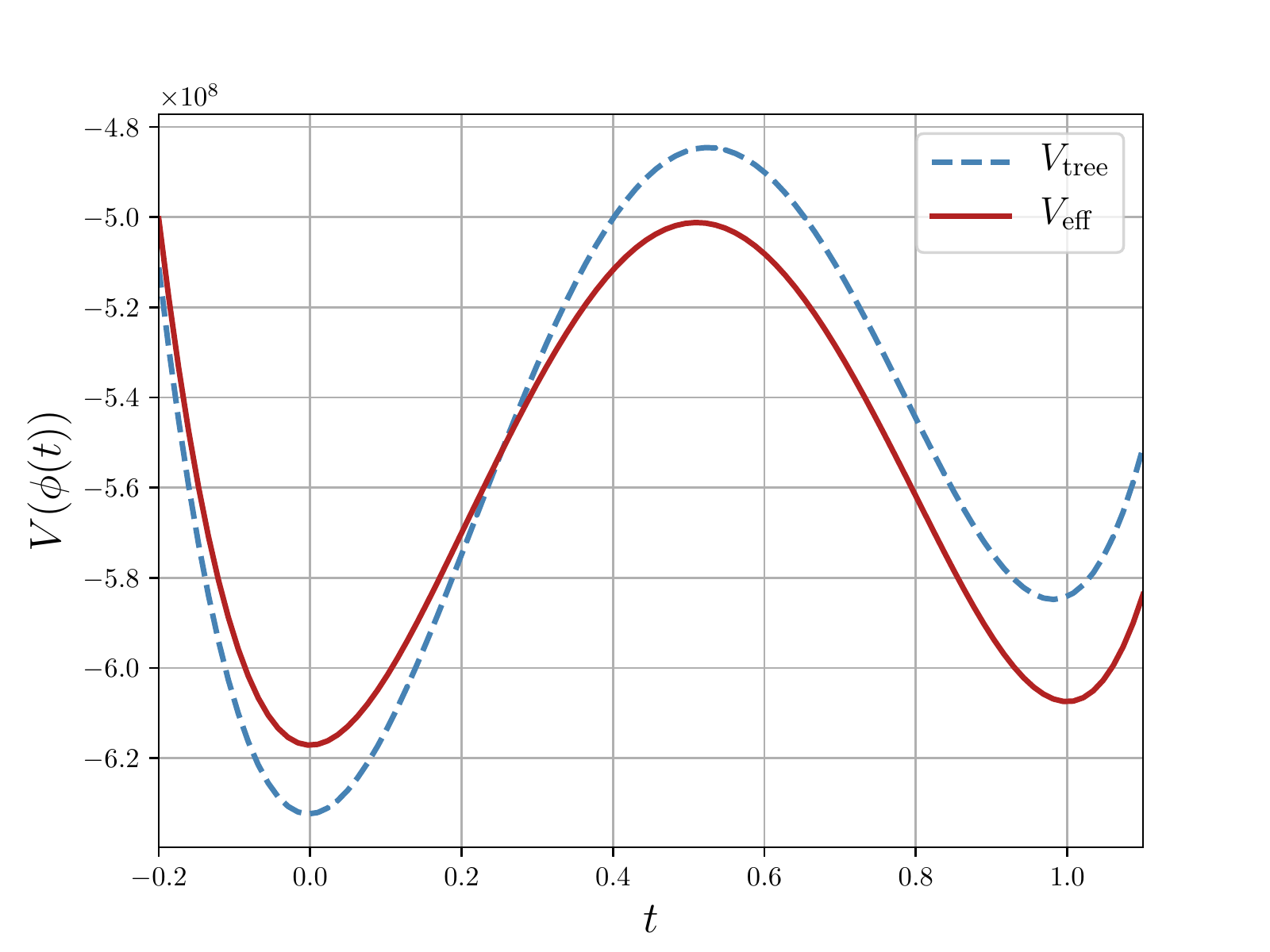}
        \caption{}
        \label{fig:pot_1D_normal}
    \end{subfigure}
    \caption{One dimensional slices of the effective scalar potential. The horizontal-axis represents vacuum configurations interpolating between the $\CB$ and $\EW$ vacuua. i.e. we interpolate between $\phi(t) = (1-t)\phi_{\EW} + t\phi_{\CB}$. Hence, at $t=0$, $\phi(t=0)=\phi_{\EW}$ and at $t=1$, $\phi(t=1)=\phi_{\CB}$. Figure (a) demonstrates a scenario where $V_{\mathrm{eff}}(\phi_{\EW}) < V_{\mathrm{eff}}(\phi_{\CB})$ and figure (b) demonstrates the scenario where $V_{\mathrm{eff}}(\phi_{\CB}) < V_{\mathrm{eff}}(\phi_{\EW})$. The values of the VEVs and parameters are given in Tab.~(\ref{tab:param_vals}).}\label{fig:pot_1D}
\end{figure}

We provide two sets of parameters yielding counterexamples to the tree-level theorem  in Tab.~(\ref{tab:param_vals}): one for the case where the $\CB$ vacuum is deeper than the $\EW$ one (left column) and one where the $\EW$ minimum is deeper than the $\CB$ one (right column.)\footnote{For more values, with more precision, see \url{https://github.com/LoganAMorrison/THDMMinimizer.jl/blob/master/data/verified_pos_mass_a1.csv} (for parameters for which $\phi_{\CB} < \phi_{\EW}$) and \url{https://github.com/LoganAMorrison/THDMMinimizer.jl/blob/master/data/verified_pos_mass_a2.csv}) (for parameters for which $\phi_{\EW} < \phi_{\CB}$.)} We reiterate that both of these points are convex at tree-level, meaning all of the squared scalar masses at tree-level are positive at the one-loop extrema, yielding no complex contributions to the effective potential. To better visualize the behaviour of the potential at both extrema for both sets of parameters, consider Fig.~(\ref{fig:pot_1D}). Of course, it is impossible to visualize the full, 8-dimensional potential at these points, but to give some sense of what it looks like in the vicinity of the one-loop vacuua, we resort to one- and two-dimensional ``slices". Thus, in Fig.~(\ref{fig:pot_1D}), we display the one-loop and tree-level potential evaluated at each vacuum and along a line linearly interpolating the $\EW$ and $\CB$ vacuua for the parameters/vacuua given in Tab.~(\ref{tab:extrema_type}). This means we are evaluating the potential along values of the fields given by, for each component of the doublets, $\vec{\phi}(t) = (1-t)\vec{\phi}_{\EW} + t\vec{\phi}_{\CB}$. Thus, at $t=0$, the potential is being evaluated at the $\EW$ vacuum and at $t=1$ at the $\CB$ vacuum. 

Fig.~(\ref{fig:pot_1D}) show that the potential always has minima at the $\EW$ and $\CB$ extrema, both at tree and one-loop level. This is, however,  deceiving -- at tree-level, the vacuum stability theorem states that if there is an EW minimum, any $\CB$ extrema will be a saddle point. However, the tree-level potential is not in fact at minima for both the $\EW$ and $\CB$ one-loop-vacuua. For both points given, the parameters give no solutions for the tree-level minimization conditions, and the one-loop $\EW$ vacuum is near the global tree-level vacuum but the one-loop $\CB$ is simply at some convex point at tree-level (but not an extremum). It would be easy to see that along some other direction(s) in field space the seeming tree-level extrema would not be minima at all. What is however clear from Fig.~(\ref{fig:pot_1D}) is, as soon as we realize that at one-loop both the $\EW$ and $\CB$ extra are minima, the tree-level vacuum stability theorem is once again violated at the one-loop level -- {\em it is possible, at one-loop, to obtain a potential with an electroweak breaking minimum, which also possesses a deeper charge breaking minimum}. Thus the absolute stability of 2HDM $\EW$ minima found at tree-level is broken by radiative corrections -- the quantum mechanical effects on the effective potential can change the vacuum properties of the model.
\begin{figure}[ht]
    \centering
    \includegraphics[width=\textwidth]{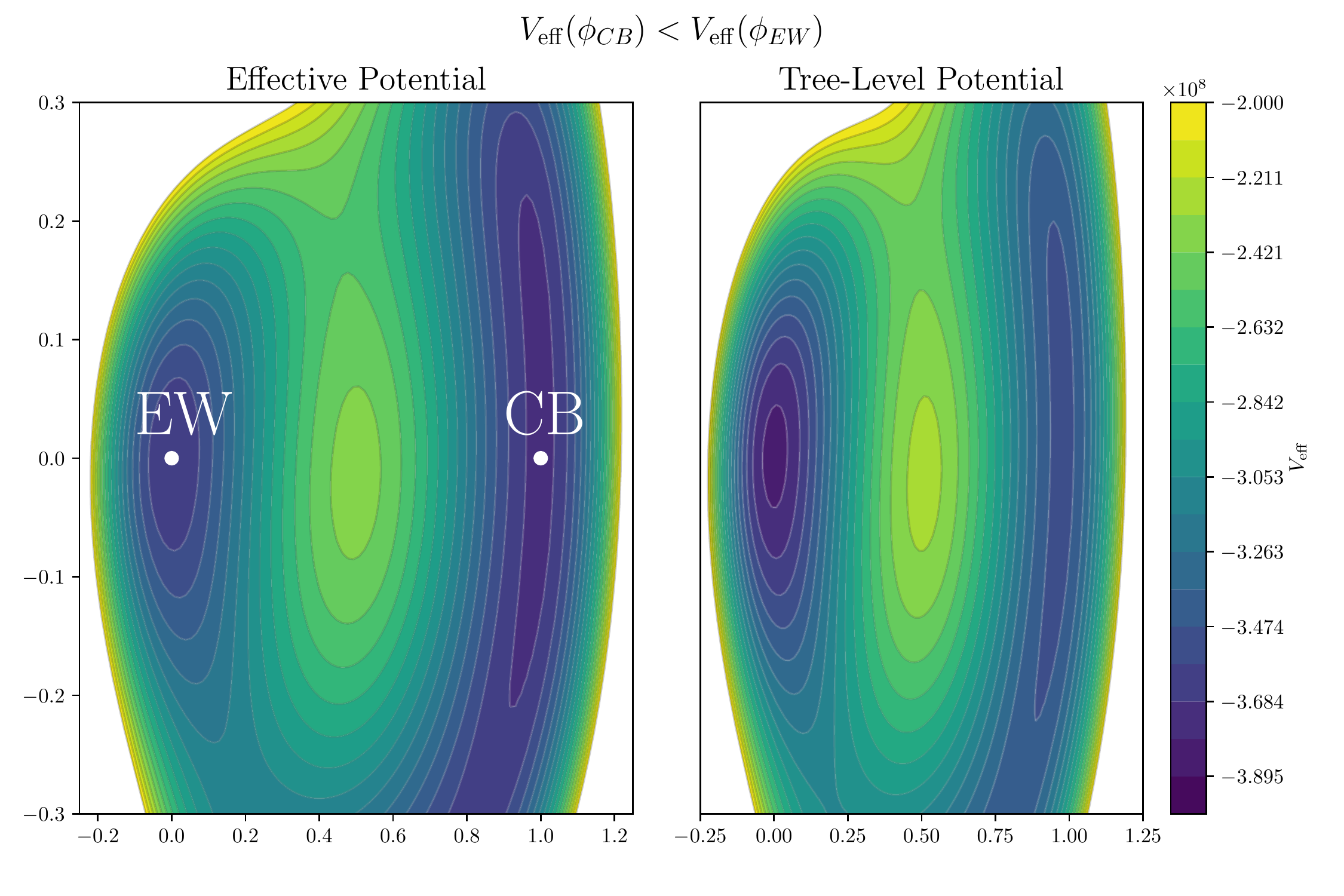}
    \caption{A two-dimensional slice of the effective potential (left) and the corresponding tree-level potential (right) for the case where $V_{\mathrm{eff}}(\phi_{\CB}) < V_{\mathrm{eff}}(\phi_{\EW})$. The horizontal axis is identical to that of Fig.~(\ref{fig:pot_1D}). The vertical axis is an line in $r_{1}-r_{2}-c_1$ space orthogonal to the $t$-axis, i.e. orthogonal to a line connecting $\vec{\phi}_{\EW}$ and $\vec{\phi}_{\CB}$. The scale of $s$ vertical axis is identical to the scale of the horizontal axis, i.e. distance in field space from $(t=0,s=0)$ and $(t=1,s=0)$ is identical to the distance in field space between $(t=0,s=0)$ and $(t=0,s=1)$ (both these distances are the distances between $\vec{\phi}_{\EW}$ and $\vec{\phi}_{CB}$.)}
    \label{fig:pot_2D_cb}
\end{figure}
\begin{figure}[ht]
    \centering
    \includegraphics[width=\textwidth]{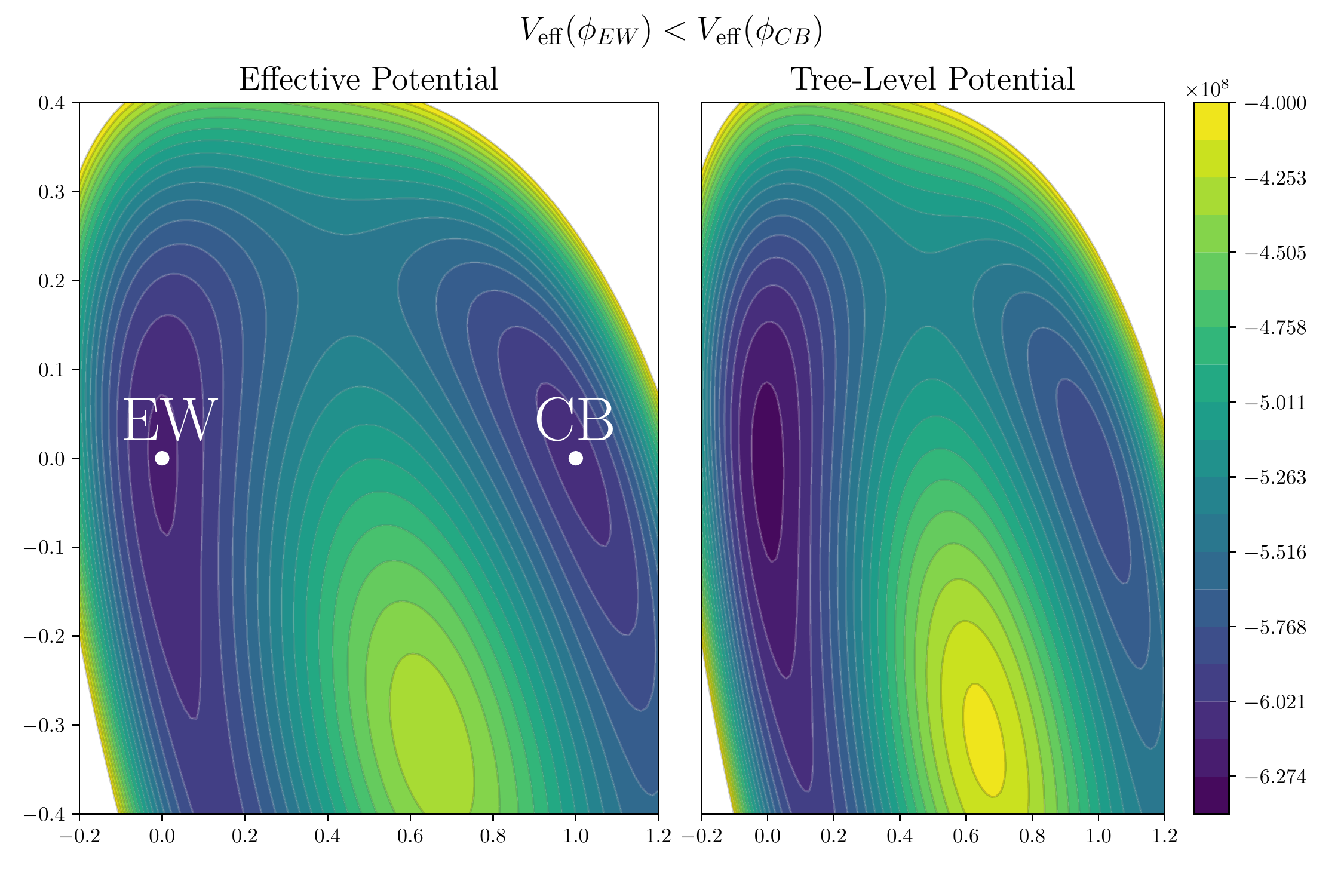}
    \caption{Same as in Fig.~(\ref{fig:pot_2D_cb}) but for the case where $V_{\mathrm{eff}}(\phi_{\EW}) < V_{\mathrm{eff}}(\phi_{\CB})$. }\label{fig:pot_2D_normal}
\end{figure}

To further illustrate the behaviour of the 2HDM potential close to these extrema consider Figs.~(\ref{fig:pot_2D_cb}) and~(\ref{fig:pot_2D_normal}). There we display a two-dimensional slice of the effective and tree-level potentials. In these figures, the horizontal axis is identical to that of the one-dimensional plots of Fig.~(\ref{fig:pot_1D}) -- that is, a line interpolating between both one-loop minima. The vertical axis in Figs.~(\ref{fig:pot_2D_cb}) and~(\ref{fig:pot_2D_normal}) represents variation along a direction $s$ in $r_1-r_2-c_1$ space which is orthogonal to the line interpolating the $\EW$ and $\CB$ vacuua. These figures give us a slightly more convincing visualization of the minimization at the $\EW$ and $\CB$ vacuua, and show the distortion induced upon the tree-level potential by the loop corrections. Having said that, they are nonetheless incomplete images of the full 8-dimensional picture and can not illustrate, for instance, the conversion between tree-level saddle points and one-loop extrema that the violation of the tree-level vacuum theorem implies. 
\begin{figure}[ht!]
    \centering
    \begin{subfigure}[b]{0.49\textwidth}
        \includegraphics[width=\textwidth]{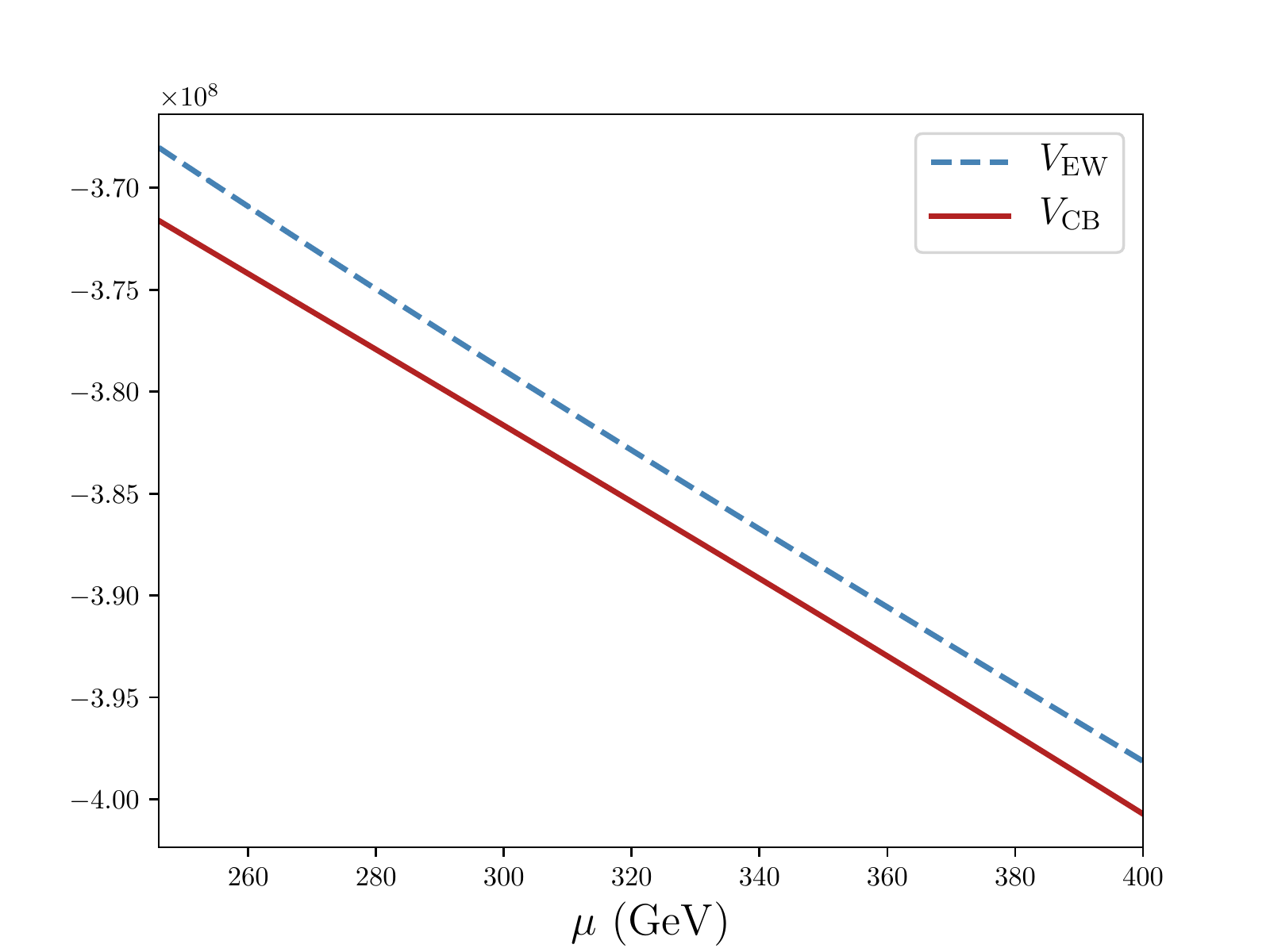}
        \caption{}\label{fig:cb_rge}
    \end{subfigure}
    \begin{subfigure}[b]{0.49\textwidth}
        \includegraphics[width=\textwidth]{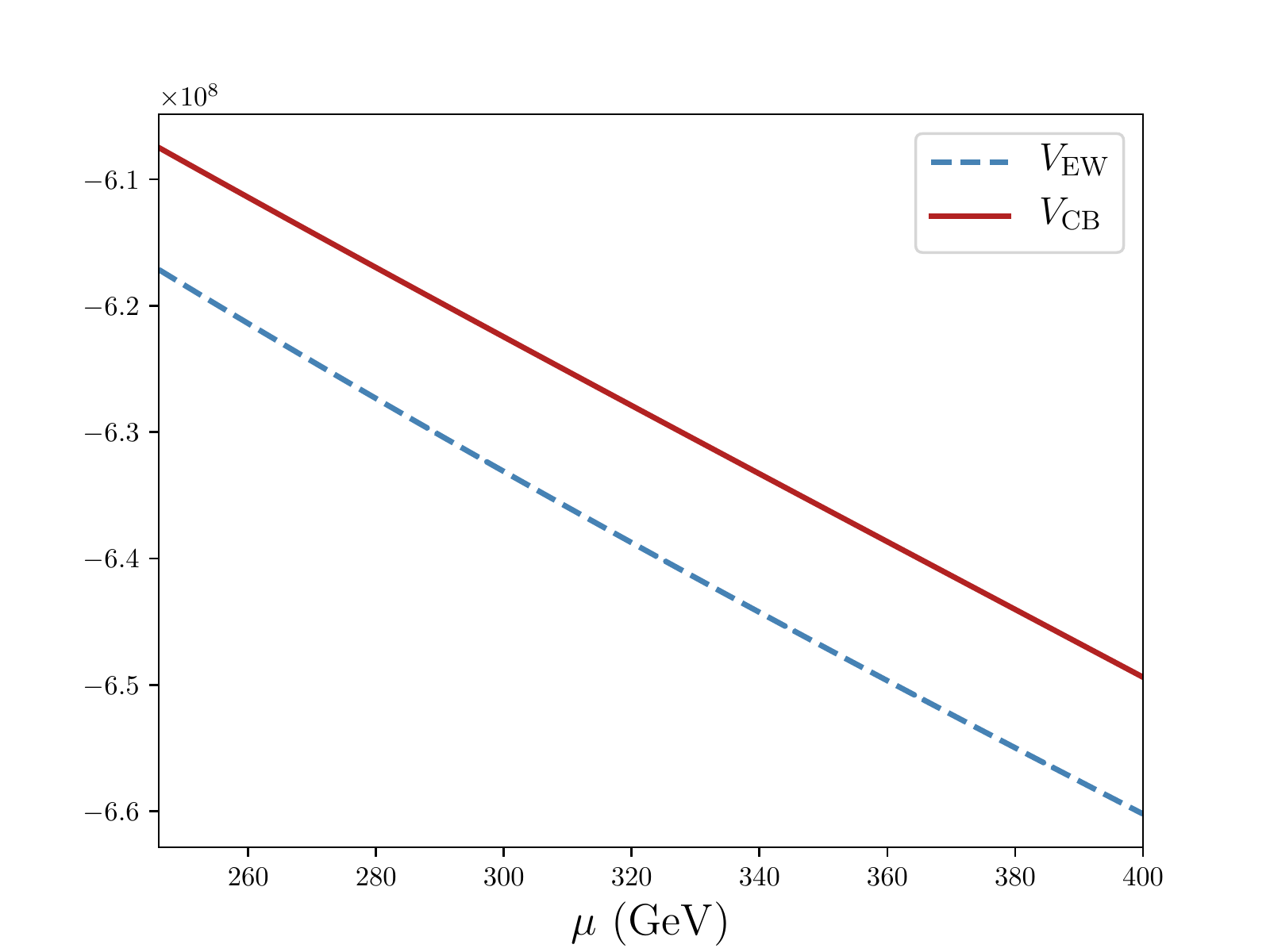}
        \caption{}
        \label{fig:normal_rge}
    \end{subfigure}
    \caption{Running of the values of the effective potential evaluated at its $\EW$ and $\CB$ minima for the case where  $V_{\mathrm{eff}}(\phi_{\CB}) < V_{\mathrm{eff}}(\phi_{\EW})$ (left) and $V_{\mathrm{eff}}(\phi_{\EW}) < V_{\mathrm{eff}}(\phi_{\CB})$ (right). These plots demonstrate that our results are independent of the particular choice of renormalization scale that we chose.}\label{fig:rge}
\end{figure}

We have fixed the renormalization scale $\mu$ to 246 GeV, and the procedure we followed should, obviously, not depend on that choice. To verify that the results we obtained are indeed not dependent on a particular choice of $\mu$, we took the parameters given in Tab.~(\ref{tab:param_vals}) and evolved them according to their RG equations (see the appendix of Ref.~\cite{basler2018high} for explicit expressions for the RG equations for the THDM parameters.) We use the \mil{DifferentialEquations.jl}\cite{rackauckas2017differentialequations} package to perform the RG evolution of the parameters from $\mu=246~\mathrm{GeV}$ to $\mu=400~\mathrm{GeV}$. At all renormalization scales between $246-400~\mathrm{GeV}$, we re-minimize the effective potential starting from both the $\EW$ and $\CB$ vacuua, determining the new VEVs at each minimum for the new values of the parameters of the potential at the new scales. We then compute the value of the one-loop effective potential at each minimum, which we show as a function of the renormalization scale in Fig.~(\ref{fig:rge}), for both sets of parameters given in Tab.~(\ref{tab:param_vals}). As we can see, the separation of the $\EW$ and $\CB$ vacuua is preserved as a function of the renormalization scale. Also, we can see that the difference of the values of the potentials is nearly a constant, which is a consequence of the fact that the effective potential at the minimum is RG independent. The values of the effective potential only change due to us not including the RG evolution of the field-independent piece of the effective potential (which is the same for both the curves). To better understand this point consider the discussion in~\cite{Ferreira:2000hg,Ferreira:2001tk}: the one-loop effective potential depends on a set of parameters $\lambda_i$, fields $\phi_j$ and renormalization scale $\mu$, and it may be written generically as
\begin{equation}
    V(\mu,\lambda_i,\phi_j)\,=\,\Omega(\mu,\lambda_i)\,+\, V_0(\lambda_i,\phi_j)\,+\,
    \hbar\,V_1(\mu,\lambda_i,\phi_j)\,+\,\order{\hbar^2}
\end{equation}
where the field-independent term $\Omega$ is the same for any extremum of the potential. The crucial insight to understand the behavior shown in Fig.~(\ref{fig:rge}) is that, unlike what one usually thinks, the sum $V_0 + \hbar\,V_1$ of the tree-level and one-loop contributions to the potential, is {\em not} RG independent. Rather, the independence of the renormalization scale on the full effective potential, $d V/d\mu = 0$, is accomplished at the one-loop level by ``compensating" the $\mu$ dependence on $V_0 + \hbar\,V_1$ with that of the $\Omega$ term~\cite{Ford:1992mv}. But since $\Omega$ does not depend on the value of the fields it will not change between the $\EW$ and $\CB$ minima, and as such it is trivial to obtain $d (V_0 + \hbar\,V_1)_{EW}/d\mu = d (V_0 + \hbar\,V_1)_{CB}/d\mu$ -- meaning, one expects that by varying the value of the renormalization scale, the value of the potential at the $\EW$ minimum evolves ``parallel" to that of the $\CB$ minimum, and that is exactly the behavior one witnesses in Fig.~(\ref{fig:rge}). Thus the conclusion is that indeed our one-loop  result is not an artifact of a specious choice of renormalization scale, but rather it is independent of the value of $\mu$. However, we emphasize that these parameter sets exemplify the best case scenario obtained in our numerical calculations, boasting nearly perfect RG evolution. Not all parameter sets we found behave as well. In particular, we find some parameter sets for which the RG curves cross. Crossing of the RG curves signals that 2-loop corrections to the effective potential and RG equations are likely important for those particular sets of parameters.

Before concluding, it is worth mentioning the consequences of these results. At tree-level, it was clear that, since it is impossible to have simultaneous $\EW$ and $\CB$ minima, an $\EW$ vacuum would be stable 
against the possibility of charge breaking, {\em i.e}. no tunneling could occur that would spoil the residual $\mathrm{U}(1)_{\mathrm{EM}}$ gauge symmetry and disastrously give the photon a mass. This is no longer the case when one considers the quantum corrections to the classical potential. That is, simultaneous $\EW$ and $\CB$ minima {\em can exist} at one-loop. This implies that, if we lived in a $\EW$ vacuum of the one-loop effective potential in a scenario where there is an additional, deeper $\CB$ vacuum, it would be possible to tunnel to the $\CB$ vacuum. The decay rate of the $\EW$ vacuum would be highly suppressed since the simultaneous minima are only realized at one-loop. In particular, we would then expect the decay rate to be a two-loop effect
and of the order $\order{\hbar^2}$.

\section{Conclusions}
\label{sec:conc}

In this work, we have analyzed the vacuum structure of the one-loop effective potential of the 2HDM. At tree-level, the 2HDM scalar potential is found to have a remarkable stability -- any minimum which breaks the ordinary electroweak symmetries and thus preserves charge conservation (and furthermore, also preserves CP) is guaranteed to be stable against the possibility of charge breaking vacuua -- meaning, any CB extremum that eventually might coexist with that minimum is guaranteed to lie above it, and furthermore to be a saddle point. This theorem was found in 2004 via analytical calculations with the 2HDM potential, along with a series of other remarkable results concerning the model's vacuum structure~\cite{Ferreira:2004yd,Barroso:2005sm,Ivanov:2006yq,Ivanov:2007de}.

The first hint that these vacuum stability theorems might not hold at one-loop was obtained analyzing the coexistence of neutral minima in a version of the 2HDM, the Inert Model. Comparing tree-level minima with one-loop ones, using the formalism of the effective potential, it was possible to show that the loop corrections might indeed change the nature of the vacuum -- for certain choices of parameters, a minimum which at tree-level was global would become a local one at one-loop~\cite{Ferreira:2015pfi}. It then became clear that an analysis of the one-loop potential was required to ascertain whether the stability of EW minima against deeper CB vacuua remained a valid conclusion. This work shows that {\em the theorem does not hold at one loop}.

We have indeed obtained, through extensive numerical scans of the parameters of the model, many cases where an EW minimum of the one-loop effective potential can coexist with a CB minimum -- this is a first violation of the tree-level theorem, which stated that an EW minimum implied necessarily CB saddle points. We have also determined that {\em one-loop EW minima can coexist with deeper CB minima} - and hence the tree-level stability against CB of the 2HDM no longer holds at one-loop. The conclusion one must draw from these results is that quantum corrections to the potential may change the vacuum structure of said potential. Conclusions drawn at tree level for which kind of minimum is the global one, and whether it is stable, may well not survive a higher-order calculation. And this, in fact, perhaps should not surprise us -- after all, this is indeed what one already obtained in the case of the Coleman-Weinberg potential~\cite{Coleman:1973jx}. 

Our calculations were performed using tried-and-true computational algorithms and numerical minimization routines which are widely available, and we offer two examples of parameter sets to be checked by interested readers. Issues of gauge dependence of the effective potential should not affect the validity of the conclusions drawn here since we are fundamentally comparing the value of the effective potential in different minima. Though we only included the contribution of the top quark, clearly the results would not qualitatively mutate with the inclusion of further fermions. And the calculations underwent a rigorous check via the computation, at each EW and CB extremum, of the respective one-loop Hessian matrices. That check had a twofold purpose: to verify the nature of any given extremum, so that we could be certain when claiming to have found minima and to verify whether the correct number of Goldstone bosons was found -- three for any EW vacuum, four for a CB one. Further, a verification of the independence of our results from the value of the renormalization scale $\mu$ we chose was undertaken -- an RG evolution of the parameters of the potential in an interval of values of $\mu$ was performed, followed by a re-minimization of the potential to obtain the values of the VEVs at each new scale. The comparison of the values of the potential showed that the relative depth of the minima remained unchanged with the renormalization scale, and thus our conclusions are RG stable.

Should this mean that we are witnessing a breakdown in perturbation theory, wherein higher-order corrections invalidate our calculations? Hardly -- the RG evolution performed showed us that perturbation theory is working as one would expect. The results of~\cite{Ferreira:2015pfi} should further illuminate our conclusions -- what was found there was that loop corrections could change tree-level expectations for the nature of the vacuua by ``swapping" global and local minima, but that this could only occur if both minima were nearly degenerate. Thus, one concludes, at least for the results of~\cite{Ferreira:2015pfi}, the loop corrections are small and acceptable perturbations that ``flip" the system between two states of nearly degenerate energies.  Likewise, the interpretation of the results we present in the current work points to perturbation theory still holding: a vast numerical scan of the model's parameters only yields counterexamples to the tree-level theorem for a small subset of the parameter space. Also, the tree-level result was strongly dependent on the specific form of the potential; of its derivatives; of the scalar squared masses. At one-loop something remarkable occurs -- the vacuum is determined, not only by the scalar sector but by all sectors of the theory, gauge and Yukawa included. It is therefore unsurprising that different statements can be made.

To conclude, the 2HDM electroweak vacuua is {\em not} guaranteed to be stable against charge breaking vacuua -- there may well be, for certain regions of parameter space, deeper CB minima below an EW one, and it may well happen that the tunneling time to the deeper minimum is smaller than the age of the universe. Though we expect this situation to be rare, this work raises the necessity to perform a wide reassessment of bounds imposed upon the parameters of the 2HDM, by fully analyzing the one-loop vacuum structure of the model. The task is not an easy one, for the one-loop effective potential is very complex and unwieldy, especially at CB vacuua. Finally, two comments-- first, we have used the results  from~\cite{Ferreira:2004yd,Barroso:2005sm} concerning the simplest form one could take for CB vacuua; but those results stem from an analysis of the tree-level potential, so they too might change when considering a one-loop calculation. Second, in~\cite{Ferreira:2004yd,Barroso:2005sm} the tree-level theorems deduced concerned the stability of EW vacuua against, not only CB, but also minima with spontaneous CP violation. As for the case of CB, the conclusion therein obtained was that any EW minimum cannot have a deeper CP breaking extremum, and any such extremum is found to lie above the EW minimum and be a saddle point. Given that we have shown that at one-loop an EW could coexist with a deeper CB minimum, there are strong reasons to believe that the same will apply to coexistence with CP minima.

\acknowledgments

PF's work is supported under CERN fund grant CERN/FIS-PAR/0002/2017, CFTC-UL grant
 UID/FIS/00618/2019 and also partially
supported by the Polish National Science Centre HARMONIA grant under contract
UMO-2015/18/M/ST2/00518 (2016-2019). LM and SP are partly supported by the U.S. Department of Energy grant number de-sc0010107.

\appendix

\section{Forward-Mode Automatic Differentiation}\label{app:autodiff}
In this appendix, we explain the technologies we use to numerically compute derivatives. There are many ways to numerically compute derivatives. The standard way is to use finite differences in which the derivative is approximated using (with forward finite differences):
\begin{align}
    f'(x) \approx \dfrac{f(x + \epsilon) - f(x)}{\epsilon} + \order{\epsilon}
\end{align}
This method suffers from many issues. First off, to get a good approximation of the derivative, one would like to make $\epsilon$ as small as possible. However, due to the finite precision of machine numbers, as $\epsilon$ becomes sufficiently small, round-off errors will seep into the calculation and the error in the approximation will increase~\cite{baydin2018automatic}. Thus, there is a given value of $\epsilon$ for which finite-differences yields the smallest error and one can do no better. Another method for evaluating derivatives is to use the complex-step method~\cite{martins_sturdza_alonso_2003}, in which the derivative is approximated using:
\begin{align}
    f'(x) \approx \Im\dfrac{f(x+i\epsilon)}{\epsilon}
\end{align}
This method doesn't suffer from the round-off errors that arise from finite differencing. $\epsilon$ can be taken arbitrarily small. However, the complex step method requires one to only use real numbers (if one mixes complex derivatives with the complex step method, the results will be non-sense.)

A slightly more complicated, but robust method of numerically computing derivatives in forward-mode automatic differentiation~\cite{baydin2018automatic}. The core idea of forward-mode automatic differentiation is the concept of dual-numbers. A dual-number is defined similarly to infinitesimals:
\begin{align}
    d &= a + \epsilon b
\end{align}
where $\epsilon$ has the property that $\epsilon^2 = 0$. The algebra of dual numbers is defined as follows:
\begin{align}
    d_{1} + d_{2} &= (a_{1} + \epsilon b_{1}) + (a_{2} + \epsilon b_{2}) \equiv (a_{1} + a_{2}) + \epsilon (b_{1} + b_{2})\\
    d_{1} \cdot d_{2} &= (a_{1} + \epsilon b_{1}) \cdot (a_{2} + \epsilon b_{2}) \equiv (a_{1}a_{2}) + \epsilon (a_{1}b_{2} + a_{2}b_{1})
\end{align}
When we evaluate a function $f(x)$ at a dual number, we obtain the standard infinitesimal shift of the function:
\begin{align}
    f(d) &= f(a+\epsilon b) \equiv f(a) + \epsilon b f'(a)
\end{align}
If we set $d = x + \epsilon$ (i.e. set $b=1$), then we find $f(x+\epsilon) = f(x) + \epsilon f'(x)$. We thus obtain $f(x)$ and $f'(x)$ by evaluating $f$ at the dual number $x + \epsilon$. Using dual-numbers provides us with a method of computing exact derivatives (up to machine precision.) Dual-numbers can also be used to compute higher-order derivatives by nesting dual-numbers: i.e. have dual-numbers of dual-numbers. For example, if $d = a + \epsilon_{1} b$ with $a = a_{1} + a_{2}\epsilon_{2}$ and $b = b_{1} + b_{2}\epsilon_{2}$, with $\epsilon_{1}^2 = \epsilon_{2}^2 = 0$ and $\epsilon_{1}\epsilon_{2}\neq0$, then we find the following:
\begin{align}
    f(d) &= f(a) + \epsilon_{1} b f'(a)\\
    &= f(a_{1}) + \epsilon_{2} a_{2}f'(a_{1}) + \epsilon_{1}(b_{1}+\epsilon_{2}b_{2})\cdot(f'(a_{1} + \epsilon_{2}f''(a_{1})))\\
    &= f(a_{1})
    +b_{1}\epsilon_{1}f'(a_{1})
    +a_{2}\epsilon_{2}f'(a_{1})
    +b_{2}\epsilon_{1}\epsilon_{2}f'(a_{1})
    +a_{2} b_{1} \epsilon_{1} \epsilon_{2}f''(a_{1})
\end{align}
If we set $a = x +\epsilon_{2}$ and $b = 1 + 0\epsilon_{2}$, then we obtain:
\begin{align}
    f((x+\epsilon_{2}) + \epsilon_{1}(1 + 0\epsilon_{2}))
    &= f(x)
    +\epsilon_{1}f'(x)
    +\epsilon_{2}f'(x)
    + \epsilon_{1} \epsilon_{2}f''(x)
\end{align}
Hence, the $\epsilon_{1}$ component of the number gives the first derivative of $f$ and the $\epsilon_{1}\epsilon_{2}$-component gives the second derivative of $f$. If we continue nesting dual-numbers, we can compute arbitrary derivatives of $f(x)$.

Given the power of template meta-programming and multiple-dispatch built into \mil{Julia}, it is an easy task to implement dual-numbers numerically. Below we provide code snippets of how this is done (note that this is not what we use, instead we use \mil{ForwardDiff.jl} a well-developed \mil{Julia} package.) The basic idea of implementing dual-numbers is to define a new type, which we call \mil{Dual}. We then overload all necessary operations that we need, i.e., addition, subtraction, multiplication, division and any other functions we wish to use with dual-numbers. Our type \mil{Dual} contains two attributes: \mil{val} (the real component of the dual-number) and \mil{eps} (the infinitesimal part):

\begin{minted}{julia}
struct Dual{T<:Real} <:Real
    val::T # real component of the dual number.
    eps::T # infinitesimal component of the dual number. eps^2 = 0
end
\end{minted}
The remaining implementation of the \mil{Dual} type is to define all the overloads of functions we want to use \mil{Dual} numbers with. For example, we can define multiplication and the trignometric \mil{sine} function as follows:
\begin{minted}{julia}
# Overload the `*` operator
function Base.:*(z::Dual{T}, w::Dual{T}) where T<:Real 
    Dual{T}(z.val * w.val, z.val * w.eps + z.eps * w.val)
end
# Overload the `sin` function
Base.sin(z::Dual{T}) where T<:Real = Dual{T}(sin(z.val), cos(z.eps))
\end{minted}
Then, one can easily perform calculations of a function and its derivative. For example, take $d_1 = 1.0 + \epsilon$ and $d_{2} = 2.0 + 0\epsilon$. Then, if we evaluate $d_{1}*d_{2}$, we find:
\begin{minted}{julia}
julia> d1 = Dual{Float64}(1.0, 1.0)
julia> d2 = Dual{Float64}(2.0, 0.0)
julia> d1 * d2
Dual{Float64}(2.0, 2.0)
\end{minted}
where the second component of the dual number is: $\frac{\partial}{\partial x} (xy) = y = 2$. Another example would be to take the sine of a dual number:
\begin{minted}{julia}
julia> sin(d1)
Dual{Float64}(0.8415, 0.5403)
\end{minted}
which we notice has $\sin(1)$ in the first component and $\cos(0)$ in the second component. Once basic operations like the above have been defined, one can then chain together very complicated functions and easily obtain their derivatives. Additionally, we can easily take second derivatives as well by nesting the dual numbers. If we define a cos overload, we can then take the second derivative of the sine function:
\begin{minted}{julia}
Base.cos(z::Dual{T}) where T<:Real = Dual{T}(cos(z.val), -sin(z.eps))
julia> d3 = Dual{Dual{Float64}}(Dual{Float64}(1., 0.), Dual{Float64}(0., 1.))
julia> sin(d3)
Dual{Dual{Float64}}(Dual{Float64}(0.8415, 1.0), Dual{Float64}(1.0, -0.8415))
\end{minted}
Here, the second component of the first dual is $d/dx \sin(x) = \cos(1)$, the first component of the second dual is the same and the second component of the second dual is the second derivative of $\sin$ at $x = 1$.



\bibliographystyle{JHEP}
\bibliography{references}

\providecommand{\href}[2]{#2}\begingroup\raggedright\begin{thebibliography}{10}

\bibitem{Aad:2012tfa}
{\bf ATLAS} Collaboration, G.~Aad et~al., {\it {Observation of a new particle
  in the search for the Standard Model Higgs boson with the ATLAS detector at
  the LHC}},  {\em Phys. Lett.} {\bf B716} (2012) 1--29,
  [\href{http://xxx.lanl.gov/abs/1207.7214}{{\tt arXiv:1207.7214}}].

\bibitem{Chatrchyan:2012xdj}
{\bf CMS} Collaboration, S.~Chatrchyan et~al., {\it {Observation of a new boson
  at a mass of 125 GeV with the CMS experiment at the LHC}},  {\em Phys. Lett.}
  {\bf B716} (2012) 30--61, [\href{http://xxx.lanl.gov/abs/1207.7235}{{\tt
  arXiv:1207.7235}}].

\bibitem{Khachatryan:2016vau}
{\bf ATLAS, CMS} Collaboration, G.~Aad et~al., {\it {Measurements of the Higgs
  boson production and decay rates and constraints on its couplings from a
  combined ATLAS and CMS analysis of the LHC $pp$ collision data at $\sqrt{s}=$
  7 and 8 TeV}},  \href{http://xxx.lanl.gov/abs/1606.0226}{{\tt
  arXiv:1606.0226}}.

\bibitem{Lee:1973iz}
T.~D. Lee, {\it {A Theory of Spontaneous T Violation}},  {\em Phys. Rev.} {\bf
  D8} (1973) 1226--1239. [,516(1973)].

\bibitem{Branco:2011iw}
G.~C. Branco, P.~M. Ferreira, L.~Lavoura, M.~N. Rebelo, M.~Sher, and J.~P.
  Silva, {\it {Theory and phenomenology of two-Higgs-doublet models}},  {\em
  Phys. Rept.} {\bf 516} (2012) 1--102,
  [\href{http://xxx.lanl.gov/abs/1106.0034}{{\tt arXiv:1106.0034}}].

\bibitem{Frere:1983ag}
J.~M. Frere, D.~R.~T. Jones, and S.~Raby, {\it {Fermion Masses and Induction of
  the Weak Scale by Supergravity}},  {\em Nucl. Phys.} {\bf B222} (1983)
  11--19.

\bibitem{Velhinho:1994np}
J.~Velhinho, R.~Santos, and A.~Barroso, {\it {Tree level vacuum stability in
  two Higgs doublet models}},  {\em Phys. Lett.} {\bf B322} (1994) 213--218.

\bibitem{Gunion:2002zf}
J.~F. Gunion and H.~E. Haber, {\it {The CP conserving two Higgs doublet model:
  The Approach to the decoupling limit}},  {\em Phys. Rev.} {\bf D67} (2003)
  075019, [\href{http://xxx.lanl.gov/abs/hep-ph/0207010}{{\tt
  hep-ph/0207010}}].

\bibitem{Ferreira:2004yd}
P.~M. Ferreira, R.~Santos, and A.~Barroso, {\it {Stability of the tree-level
  vacuum in two Higgs doublet models against charge or CP spontaneous
  violation}},  {\em Phys. Lett.} {\bf B603} (2004) 219--229,
  [\href{http://xxx.lanl.gov/abs/hep-ph/0406231}{{\tt hep-ph/0406231}}].
  [Erratum: Phys. Lett.B629,114(2005)].

\bibitem{Barroso:2005sm}
A.~Barroso, P.~M. Ferreira, and R.~Santos, {\it {Charge and CP symmetry
  breaking in two Higgs doublet models}},  {\em Phys. Lett.} {\bf B632} (2006)
  684--687, [\href{http://xxx.lanl.gov/abs/hep-ph/0507224}{{\tt
  hep-ph/0507224}}].

\bibitem{Maniatis:2006fs}
M.~Maniatis, A.~von Manteuffel, O.~Nachtmann, and F.~Nagel, {\it {Stability and
  symmetry breaking in the general two-Higgs-doublet model}},  {\em Eur. Phys.
  J.} {\bf C48} (2006) 805--823,
  [\href{http://xxx.lanl.gov/abs/hep-ph/0605184}{{\tt hep-ph/0605184}}].

\bibitem{Nishi:2006tg}
C.~C. Nishi, {\it {CP violation conditions in N-Higgs-doublet potentials}},
  {\em Phys. Rev.} {\bf D74} (2006) 036003,
  [\href{http://xxx.lanl.gov/abs/hep-ph/0605153}{{\tt hep-ph/0605153}}].
  [Erratum: Phys. Rev.D76,119901(2007)].

\bibitem{Ivanov:2006yq}
I.~P. Ivanov, {\it {Minkowski space structure of the Higgs potential in 2HDM}},
   {\em Phys. Rev.} {\bf D75} (2007) 035001,
  [\href{http://xxx.lanl.gov/abs/hep-ph/0609018}{{\tt hep-ph/0609018}}].
  [Erratum: Phys. Rev.D76,039902(2007)].

\bibitem{Ivanov:2007de}
I.~P. Ivanov, {\it {Minkowski space structure of the Higgs potential in 2HDM.
  II. Minima, symmetries, and topology}},  {\em Phys. Rev.} {\bf D77} (2008)
  015017, [\href{http://xxx.lanl.gov/abs/0710.3490}{{\tt arXiv:0710.3490}}].

\bibitem{Maniatis:2007vn}
M.~Maniatis, A.~von Manteuffel, and O.~Nachtmann, {\it {CP violation in the
  general two-Higgs-doublet model: A Geometric view}},  {\em Eur. Phys. J.}
  {\bf C57} (2008) 719--738, [\href{http://xxx.lanl.gov/abs/0707.3344}{{\tt
  arXiv:0707.3344}}].

\bibitem{Barroso:2007rr}
A.~Barroso, P.~M. Ferreira, and R.~Santos, {\it {Neutral minima in two-Higgs
  doublet models}},  {\em Phys. Lett.} {\bf B652} (2007) 181--193,
  [\href{http://xxx.lanl.gov/abs/hep-ph/0702098}{{\tt hep-ph/0702098}}].

\bibitem{Deshpande:1977rw}
N.~G. Deshpande and E.~Ma, {\it {Pattern of Symmetry Breaking with Two Higgs
  Doublets}},  {\em Phys. Rev.} {\bf D18} (1978) 2574.

\bibitem{Barbieri:2006dq}
R.~Barbieri, L.~J. Hall, and V.~S. Rychkov, {\it {Improved naturalness with a
  heavy Higgs: An Alternative road to LHC physics}},  {\em Phys. Rev.} {\bf
  D74} (2006) 015007, [\href{http://xxx.lanl.gov/abs/hep-ph/0603188}{{\tt
  hep-ph/0603188}}].

\bibitem{Cao:2007rm}
Q.-H. Cao, E.~Ma, and G.~Rajasekaran, {\it {Observing the Dark Scalar Doublet
  and its Impact on the Standard-Model Higgs Boson at Colliders}},  {\em Phys.
  Rev.} {\bf D76} (2007) 095011, [\href{http://xxx.lanl.gov/abs/0708.2939}{{\tt
  arXiv:0708.2939}}].

\bibitem{Ferreira:2015pfi}
P.~M. Ferreira and B.~Swiezewska, {\it {One-loop contributions to neutral
  minima in the inert doublet model}},  {\em JHEP} {\bf 04} (2016) 099,
  [\href{http://xxx.lanl.gov/abs/1511.0287}{{\tt arXiv:1511.0287}}].

\bibitem{Davidson:2005cw}
S.~Davidson and H.~E. Haber, {\it {Basis-independent methods for the
  two-Higgs-doublet model}},  {\em Phys. Rev.} {\bf D72} (2005) 035004,
  [\href{http://xxx.lanl.gov/abs/hep-ph/0504050}{{\tt hep-ph/0504050}}].
  [Erratum: Phys. Rev.D72,099902(2005)].

\bibitem{Glashow:1976nt}
S.~L. Glashow and S.~Weinberg, {\it {Natural Conservation Laws for Neutral
  Currents}},  {\em Phys. Rev.} {\bf D15} (1977) 1958.

\bibitem{Paschos:1976ay}
E.~A. Paschos, {\it {Diagonal Neutral Currents}},  {\em Phys. Rev.} {\bf D15}
  (1977) 1966.

\bibitem{Pilaftsis:1999qt}
A.~Pilaftsis and C.~E.~M. Wagner, {\it {Higgs bosons in the minimal
  supersymmetric standard model with explicit CP violation}},  {\em Nucl.
  Phys.} {\bf B553} (1999) 3--42,
  [\href{http://xxx.lanl.gov/abs/hep-ph/9902371}{{\tt hep-ph/9902371}}].

\bibitem{Ginzburg:2002wt}
I.~F. Ginzburg, M.~Krawczyk, and P.~Osland, {\it {Two Higgs doublet models with
  CP violation}},  in {\em {Linear colliders. Proceedings, International
  Workshop on physics and experiments with future electron-positron linear
  colliders, LCWS 2002, Seogwipo, Jeju Island, Korea, August 26-30, 2002}},
  pp.~703--706, 2002.
\newblock \href{http://xxx.lanl.gov/abs/hep-ph/0211371}{{\tt hep-ph/0211371}}.
\newblock [,703(2002)].

\bibitem{Khater:2003wq}
W.~Khater and P.~Osland, {\it {CP violation in top quark production at the LHC
  and two Higgs doublet models}},  {\em Nucl. Phys.} {\bf B661} (2003)
  209--234, [\href{http://xxx.lanl.gov/abs/hep-ph/0302004}{{\tt
  hep-ph/0302004}}].

\bibitem{ElKaffas:2007rq}
A.~W. El~Kaffas, P.~Osland, and O.~M. Ogreid, {\it {CP violation, stability and
  unitarity of the two Higgs doublet model}},  {\em Nonlin. Phenom. Complex
  Syst.} {\bf 10} (2007) 347--357,
  [\href{http://xxx.lanl.gov/abs/hep-ph/0702097}{{\tt hep-ph/0702097}}].

\bibitem{ElKaffas:2006gdt}
A.~W. El~Kaffas, W.~Khater, O.~M. Ogreid, and P.~Osland, {\it {Consistency of
  the two Higgs doublet model and CP violation in top production at the LHC}},
  {\em Nucl. Phys.} {\bf B775} (2007) 45--77,
  [\href{http://xxx.lanl.gov/abs/hep-ph/0605142}{{\tt hep-ph/0605142}}].

\bibitem{WahabElKaffas:2007xd}
A.~Wahab El~Kaffas, P.~Osland, and O.~M. Ogreid, {\it {Constraining the
  Two-Higgs-Doublet-Model parameter space}},  {\em Phys. Rev.} {\bf D76} (2007)
  095001, [\href{http://xxx.lanl.gov/abs/0706.2997}{{\tt arXiv:0706.2997}}].

\bibitem{Osland:2008aw}
P.~Osland, P.~N. Pandita, and L.~Selbuz, {\it {Trilinear Higgs couplings in the
  two Higgs doublet model with CP violation}},  {\em Phys. Rev.} {\bf D78}
  (2008) 015003, [\href{http://xxx.lanl.gov/abs/0802.0060}{{\tt
  arXiv:0802.0060}}].

\bibitem{Grzadkowski:2009iz}
B.~Grzadkowski and P.~Osland, {\it {Tempered Two-Higgs-Doublet Model}},  {\em
  Phys. Rev.} {\bf D82} (2010) 125026,
  [\href{http://xxx.lanl.gov/abs/0910.4068}{{\tt arXiv:0910.4068}}].

\bibitem{Arhrib:2010ju}
A.~Arhrib, E.~Christova, H.~Eberl, and E.~Ginina, {\it {CP violation in charged
  Higgs production and decays in the Complex Two Higgs Doublet Model}},  {\em
  JHEP} {\bf 04} (2011) 089, [\href{http://xxx.lanl.gov/abs/1011.6560}{{\tt
  arXiv:1011.6560}}].

\bibitem{Barroso:2012wz}
A.~Barroso, P.~M. Ferreira, R.~Santos, and J.~P. Silva, {\it {Probing the
  scalar-pseudoscalar mixing in the 125 GeV Higgs particle with current data}},
   {\em Phys. Rev.} {\bf D86} (2012) 015022,
  [\href{http://xxx.lanl.gov/abs/1205.4247}{{\tt arXiv:1205.4247}}].

\bibitem{Ferreira:2011xc}
P.~M. Ferreira, L.~Lavoura, J.~P. Silva, and L.~Lavoura, {\it {A Soft origin
  for CKM-type CP violation}},  {\em Phys. Lett.} {\bf B704} (2011) 179--188,
  [\href{http://xxx.lanl.gov/abs/1102.0784}{{\tt arXiv:1102.0784}}].

\bibitem{Ferreira:2019aps}
P.~M. Ferreira and L.~Lavoura, {\it {No strong $CP$ violation up to the
  one-loop level in a two-Higgs-doublet model}},
  \href{http://xxx.lanl.gov/abs/1904.0843}{{\tt arXiv:1904.0843}}.

\bibitem{martin2002two}
S.~P. Martin, {\it Two-loop effective potential for a general renormalizable
  theory and softly broken supersymmetry},  {\em Physical Review D} {\bf 65}
  (2002), no.~11 116003.

\bibitem{Nielsen:1975fs}
N.~K. Nielsen, {\it {On the Gauge Dependence of Spontaneous Symmetry Breaking
  in Gauge Theories}},  {\em Nucl. Phys.} {\bf B101} (1975) 173--188.

\bibitem{Patel:2011th}
H.~H. Patel and M.~J. Ramsey-Musolf, {\it {Baryon Washout, Electroweak Phase
  Transition, and Perturbation Theory}},  {\em JHEP} {\bf 07} (2011) 029,
  [\href{http://xxx.lanl.gov/abs/1101.4665}{{\tt arXiv:1101.4665}}].

\bibitem{Elias-Miro:2014pca}
J.~Elias-Miro, J.~R. Espinosa, and T.~Konstandin, {\it {Taming Infrared
  Divergences in the Effective Potential}},  {\em JHEP} {\bf 08} (2014) 034,
  [\href{http://xxx.lanl.gov/abs/1406.2652}{{\tt arXiv:1406.2652}}].

\bibitem{Martin:2014bca}
S.~P. Martin, {\it {Taming the Goldstone contributions to the effective
  potential}},  {\em Phys. Rev.} {\bf D90} (2014), no.~1 016013,
  [\href{http://xxx.lanl.gov/abs/1406.2355}{{\tt arXiv:1406.2355}}].

\bibitem{Ekstedt:2018ftj}
A.~Ekstedt and J.~Löfgren, {\it {On the relationship between gauge dependence
  and IR divergences in the $\hbar$-expansion of the effective potential}},
  {\em JHEP} {\bf 01} (2019) 226,
  [\href{http://xxx.lanl.gov/abs/1810.0141}{{\tt arXiv:1810.0141}}].

\bibitem{Andreassen:2014eha}
A.~Andreassen, W.~Frost, and M.~D. Schwartz, {\it {Consistent Use of Effective
  Potentials}},  {\em Phys. Rev.} {\bf D91} (2015), no.~1 016009,
  [\href{http://xxx.lanl.gov/abs/1408.0287}{{\tt arXiv:1408.0287}}].

\bibitem{nlsolve}
K.~Carlsson, P.~K. Mogensen, S.~Villemot, S.~Lyon, M.~Gomez, C.~Rackauckas,
  T.~Holy, T.~Kelman, D.~Widmann, M.~R.~G. Macedo, and et~al., {\it
  Julianlsolvers/nlsolve.jl: v4.1.0}, .

\bibitem{RevelsLubinPapamarkou2016}
J.~{Revels}, M.~{Lubin}, and T.~{Papamarkou}, {\it Forward-mode automatic
  differentiation in julia},  {\em arXiv:1607.07892 [cs.MS]} (2016).

\bibitem{Branchina:2018qlf}
V.~Branchina, F.~Contino, and P.~M. Ferreira, {\it {Electroweak vacuum lifetime
  in two Higgs doublet models}},  {\em JHEP} {\bf 11} (2018) 107,
  [\href{http://xxx.lanl.gov/abs/1807.1080}{{\tt arXiv:1807.1080}}].

\bibitem{mogensen2018optim}
P.~K. Mogensen and A.~N. Riseth, {\it Optim: A mathematical optimization
  package for {Julia}},  {\em Journal of Open Source Software} {\bf 3} (2018),
  no.~24 615.

\bibitem{Barroso:2012mj}
A.~Barroso, P.~M. Ferreira, I.~P. Ivanov, R.~Santos, and J.~P. Silva, {\it
  {Evading death by vacuum}},  {\em Eur. Phys. J.} {\bf C73} (2013) 2537,
  [\href{http://xxx.lanl.gov/abs/1211.6119}{{\tt arXiv:1211.6119}}].

\bibitem{Barroso:2013awa}
A.~Barroso, P.~M. Ferreira, I.~P. Ivanov, and R.~Santos, {\it {Metastability
  bounds on the two Higgs doublet model}},  {\em JHEP} {\bf 06} (2013) 045,
  [\href{http://xxx.lanl.gov/abs/1303.5098}{{\tt arXiv:1303.5098}}].

\bibitem{basler2018high}
P.~Basler, P.~M. Ferreira, M.~M{\"u}hlleitner, and R.~Santos, {\it High scale
  impact in alignment and decoupling in two-higgs-doublet models},  {\em
  Physical Review D} {\bf 97} (2018), no.~9 095024.

\bibitem{rackauckas2017differentialequations}
C.~Rackauckas and Q.~Nie, {\it Differentialequations. jl--a performant and
  feature-rich ecosystem for solving differential equations in julia},  {\em
  Journal of Open Research Software} {\bf 5} (2017), no.~1.

\bibitem{Ferreira:2000hg}
P.~M. Ferreira, {\it {A Full one loop charge and color breaking effective
  potential}},  {\em Phys. Lett.} {\bf B509} (2001) 120--130,
  [\href{http://xxx.lanl.gov/abs/hep-ph/0008115}{{\tt hep-ph/0008115}}].
  [Erratum: Phys. Lett.B518,333(2001)].

\bibitem{Ferreira:2001tk}
P.~M. Ferreira, {\it {Minimization of a one loop charge breaking effective
  potential}},  {\em Phys. Lett.} {\bf B512} (2001) 379--391,
  [\href{http://xxx.lanl.gov/abs/hep-ph/0102141}{{\tt hep-ph/0102141}}].
  [Erratum: Phys. Lett.B518,334(2001)].

\bibitem{Ford:1992mv}
C.~Ford, D.~R.~T. Jones, P.~W. Stephenson, and M.~B. Einhorn, {\it {The
  Effective potential and the renormalization group}},  {\em Nucl. Phys.} {\bf
  B395} (1993) 17--34, [\href{http://xxx.lanl.gov/abs/hep-lat/9210033}{{\tt
  hep-lat/9210033}}].

\bibitem{Coleman:1973jx}
S.~R. Coleman and E.~J. Weinberg, {\it {Radiative Corrections as the Origin of
  Spontaneous Symmetry Breaking}},  {\em Phys. Rev.} {\bf D7} (1973)
  1888--1910.

\bibitem{baydin2018automatic}
A.~G. Baydin, B.~A. Pearlmutter, A.~A. Radul, and J.~M. Siskind, {\it Automatic
  differentiation in machine learning: a survey},  {\em Journal of machine
  learning research} {\bf 18} (2018), no.~153.

\bibitem{martins_sturdza_alonso_2003}
J.~R. R.~A. Martins, P.~Sturdza, and J.~J. Alonso, {\it The complex-step
  derivative approximation},  {\em ACM Transactions on Mathematical Software}
  {\bf 29} (2003), no.~3 245–262.

\end{thebibliography}\endgroup

\end{document}